\documentclass[10pt]{article}

\usepackage{amsmath, amsfonts, amssymb}
\usepackage[ansinew]{inputenc}
\usepackage[T1]{fontenc}
\usepackage{dsfont}
\usepackage{graphicx}
\usepackage[longnamesfirst,numbers,square]{natbib}
\usepackage{subfigure}
\usepackage{palatino, euler}
\usepackage[pdfpagemode=UseNone, pdfstartview={XYZ null null null}, citebordercolor={0 0.75 0}, urlbordercolor={0 0 0.75}]{hyperref} 

\title{\huge Geometric Exponents\\ of Dilute Logarithmic Minimal Models}
\author{
 {\Large Guillaume Provencher}\thanks{\tt provench@dms.umontreal.ca} \\[2mm] \textit{Département de physique} \\ \textit{Université de Montréal, C.P. 6128, succ. centre-ville} \\ \textit{Montréal, QC, Canada, H3C 3J7}\\[2mm]
{\Large Yvan Saint-Aubin}\thanks{\tt saint@dms.umontreal.ca} \\[2mm] \textit{Département de mathématiques et de statistique} \\ \textit{Université de Montréal, C.P. 6128, succ. centre-ville} \\ \textit{Montréal, QC, Canada, H3C 3J7}
 \and
 {\Large Paul A. Pearce}\thanks{\tt P.Pearce@ms.unimelb.edu.au}, \quad{\Large J\o{}rgen Rasmussen}\thanks{\tt J.Rasmussen@ms.unimelb.edu.au} \\[2mm] \textit{Department of Mathematics and Statistics} \\ \textit{University of Melbourne} \\ \textit{Parkville, Victoria 3010, Australia}
 }
 \date{\today}

\begin{document}

\maketitle  
\begin{abstract}
\noindent 
The fractal dimensions of the hull, the external perimeter and of the red bonds are measured 
through Monte Carlo simulations for dilute minimal models, and compared with 
predictions from conformal field theory and SLE methods.
The dilute models used are those first introduced by Nienhuis.
Their loop fugacity is $\beta=-2 \cos(\pi/\bar\kappa)$ where the parameter $\bar\kappa$ is linked 
to their description through conformal loop ensembles. 
It is also linked to conformal field theories through their central charges 
$c(\bar\kappa)=13-6(\bar\kappa+\bar\kappa^{-1})$ and, for the minimal models of interest 
here, $\bar\kappa=p/p'$ where $p$ and $p'$ are two coprime integers. The geometric exponents 
of the hull and external perimeter are studied for the pairs 
$(p,p')=(1,1),(2,3),(3,4),(4,5),(5,6),(5,7)$, and that of the red bonds for $(p,p')=(3,4)$. Monte Carlo 
upgrades are proposed for these models as well as several techniques to 
improve their speeds. The measured fractal dimensions are obtained by extrapolation on the 
lattice size $H,V\rightarrow\infty$. The extrapolating curves have large slopes; 
despite these, the measured dimensions coincide with theoretical predictions up to three or four 
digits. In some cases, the theoretical values lie slightly outside the confidence 
intervals; explanations of these small discrepancies are proposed.
\\[4mm]
\end{abstract}
\clearpage

\tableofcontents
\newpage


\baselineskip=18pt

\section{Introduction}

Geometric objects remain a central tool in the study of the critical behavior of statistical lattice 
models. Some of the most natural ones are the hull of a spin cluster, its mass, 
external perimeter and red bonds. Even though they were investigated as early as in the 1970's, 
their role has remained central over the years. In the 1980's, their close ties 
to conformal weights of the Virasoro algebra, and more generally with conformal field theory 
(CFT), was revealed starting with works by \citet{Saleur1987}. And in the late 
1990's, techniques from probability theory related their properties to that of random curves grown 
through stochastic Loewner evolution (SLE). The goal of this paper is to 
measure, using Monte Carlo simulations, the fractal dimensions of these objects for dilute lattice 
models. 

Recently, \citet{Saint-Aubin2009} measured these dimensions for a 
family of lattice loop models whose continuum scaling
limit is called the {\em logarithmic minimal models} 
(\citet{PRZ2006}). Their results gave compelling evidence for the theoretical 
predictions of \citet{Saleur1987} and others, and confirmed the rigorous result by 
\citet{Beffara2008} for the hull fractal dimension. Their work probed the dense phase of loop 
models and the present paper can be seen as completing their work by considering the dilute 
phase. 

We shall do so on another family of loop models based upon the celebrated 
$\mathcal{O}(n)$ model. Writing the loop fugacity as $n= -2\cos\tfrac{\pi}{\bar\kappa}$, these
loop models are well defined for all real $\bar\kappa$ values, and our methods apply 
for arbitrary values of $\bar\kappa$. Focus here is on rational values, though, 
for which the fractal dimensions are rational and expressible in terms
of conformal dimensions from an extended Kac table. We furthermore
believe that, for $\bar\kappa$ rational, these loop models converge to logarithmic CFTs in the 
continuum scaling limit and henceforth refer to them as \emph{dilute logarithmic minimal models}.

Beside the intrinsic value of checking theoretical predictions through experiments or, in the 
present case, Monte Carlo simulations, such checks often lead to improvements in 
the techniques of (numerical) experimentation. Together with Potts models, the XXZ Hamiltonian 
and other spin lattice models, dilute loop models are of great theoretical value. 
But, because the Boltzmann weights of loop models require the counting of the number of loops in 
a configuration, a task that is highly non-local, simulations of these loop models 
remain difficult. The classical algorithms, like that of \citet{Swendsen1987}, usually do not apply to 
them. There has been progress to extend these cluster algorithms to larger 
families of models, e.g.~by \citet{Chayes1998} and \citet{Deng2007} but, unfortunately, some loop 
models remain without efficient algorithms. One of the outcomes of the 
present paper is the proposed upgrade algorithm and its variants that curtail significantly the 
difficulties of visiting large non-local objects, like loops whose size is commensurate 
to that of the lattice.

This paper is organized as follows. The next section recalls the definition of the $\mathcal{O}(n)$ 
model and characterizes the interval of the parameter that will lead to their 
dilute phase. The definition of the geometric fractal dimensions for the hull, external perimeter and 
red bonds are given with their theoretical predictions. Section~3 describes 
difficulties inherent to measuring geometric exponents on dilute models: the area associated to 
the various states of boxes, the important boundary effects and the distribution 
of the defect's winding number. Section~4 presents the results and discusses some of their 
shortcomings, while section~5 contains some concluding remarks. 
Technical details are gathered in the appendices: the Monte Carlo 
algorithms are described in appendix~A, and the statistical analysis in appendix~B. Appendix~C 
recalls the definition of the dense loop models used in \citep{Saint-Aubin2009}.


\section{The $\mathcal{O}(n)$ model and dilute logarithmic minimal models}

\subsection{Loop representation of the two-dimensional $\mathcal{O}(n)$ model}

The two-dimensional $\mathcal{O}(n)$ model is a lattice spin model, whose $n$-component 
spins $s_i^\mu$, for $\mu\in\{1,2,\cdots,n\}$, are located at each site 
$i\in\mathcal{S}$ of the lattice, with $\mathcal{S}$ being the set of sites. The spins are constrained 
to live on an \mbox{$(n-1)$-dimensional} sphere of radius $\sqrt n$, i.e. they satisfy 
$\mathbf{s_i}\cdot\mathbf{s_i}=\sum_{\mu=1}^ns_i^\mu s_i^\mu=s_i^\mu s_i^\mu=n$.
As indicated, no summation is implied on Latin letters.

\subsubsection{On the honeycomb lattice}

The partition function of this model on a domain of the honeycomb lattice is defined as 
$Z=\mathsf{Tr}\left[\prod_{\langle ij\rangle}\mathrm{e}^{xs_i^\mu s_j^\mu}\right]$ 
where the product is taken over all lattice edges, written $\langle ij\rangle$, and $x$ is the 
inverse temperature. (See for example \citet{Dubail2009}.)
The high-temperature expansion, that is for $x$ small, is given by
\begin{equation}
 \label{eq:OnHoney}
 Z=\mathsf{Tr}\biggl[\prod_{\langle ij\rangle}\bigl(1+xs_i^\mu s_j^\mu\bigr)\biggr].
\end{equation}
The trace operator
\begin{equation}
 \label{eq:Trace}
 \mathsf{Tr}\bigl[\Lambda(\mathbf{s_{i_1}},\mathbf{s_{i_2}},\dotsc)\bigr]=C\left\{\prod_{i\in\mathcal{S}}\int_{\mathds{R}^n}\delta\bigl(s_i^\mu s_i^\mu-n\bigr)\,\mathrm{d}\mathbf{s_i}\right\}\Lambda(\mathbf{s_{i_1}},\mathbf{s_{i_2}},\dotsc),
\end{equation}
with $C$ a normalization factor, allows to compute the expectation value of an arbitrary function 
$\Lambda$ of the spins. It may be normalized such that the following properties hold:
\begin{align}
 \label{eq:TrProp}
 \notag\mathsf{Tr}[1] &=1																				\\
 \mathsf{Tr}\left[s_i^\mu s_j^\nu\right] &=\delta_{ij}\delta^{\mu\nu}													\\
 \notag\mathsf{Tr}\left[s_i^\mu\right] &= \mathsf{Tr}\left[\bigl(s_i^\mu\bigr)^3\right]=\mathsf{Tr}\left[\bigl(s_i^\mu\bigr)^5\right]=\dotsb=0.
\end{align}
In this case, the normalization factor is given by 
$C=\Bigl[\frac{n\pi}{2\Gamma\left(\frac{n}{2}+1\right)}\Bigr]^{-|\mathcal{S}|}$ with $|\mathcal{S}|$ 
the number of sites in 
the domain. With these properties, the partition function \eqref{eq:OnHoney} becomes a sum on all 
configurations with non-intersecting loops. This is 
so because only cyclic terms of the form 
$x^ks_{i_1}^{\mu_1}s_{i_2}^{\mu_1}s_{i_2}^{\mu_2}s_{i_3}^{\mu_2}\dotsm s_{i_k}^{\mu_k}s_{i_1}^{\mu_k}$ 
survive the trace, yielding a weight $nx^k$ for the configuration. Applying these rules leads to the 
celebrated loop partition function
\begin{equation}
 \label{eq:OnHoneyLoop}
 Z=\sum_{\mathcal{L}}x^Xn^N
\end{equation}
for the $\mathcal{O}(n)$ model on the honeycomb lattice. The sum is taken over all lattice configurations $\mathcal{L}$ of non-intersecting loops. Here $X$ 
and $N$ are respectively the total number of monomers (bonds) and the total number of loops of 
the configuration. The parameters $x$ and $n$ play 
accordingly the roles of bond and loop fugacity. The loop representation \eqref{eq:OnHoneyLoop} 
describes a larger family of models than the spin representation 
\eqref{eq:OnHoney} because $n$ may take \emph{real} values, not only positive integer ones.

\subsubsection{On the square lattice}\label{sec:squareLattice}

The trace operator \eqref{eq:Trace} forbids the presence of terms of odd powers in $s_i^\mu$, 
so diagrams of intersecting or open loops are impossible in any 
$\mathcal{L}$ on the honeycomb lattice. But loop intersections may occur in the $\mathcal{O}(n)$
model \eqref{eq:OnHoney} on the square lattice, because terms like $x^4(s_i^\mu)^4$ 
are possible there and non-zero in general. The square-lattice $\mathcal{O}(n)$ model we are 
interested in is not equivalent to \eqref{eq:OnHoney}, although it is quite similar, 
and was first defined by \citet{Nienhuis1990}. 
This model is defined on the square lattice, possesses a partition
function similar to \eqref{eq:OnHoneyLoop} and has no intersecting loops.

The spins of the model are located on the edges of the square lattice and there is now three 
interaction constants: $u$, $v$, and $w$. These parameters are understood to 
include the inverse temperature. As for the previous model, the spins are constrained by 
$s_i^\mu s_i^\mu=n$ and the partition function is defined as
\begin{equation}
 \label{eq:OnSquare}
 Z=\mathsf{Tr}\biggl[\prod_{\langle i,j,k,l\rangle}Q(\mathbf{s_i},\mathbf{s_j},\mathbf{s_k},\mathbf{s_l})\biggr]
\end{equation}
where the product is taken over all lattice faces $\langle i,j,k,l\rangle$ with $i,j,k$ and $l$ the spins 
surrounding the face in fixed order (say clockwise). The Boltzmann weight 
$Q$ is of the form
$$
 Q(\mathbf{s_i},\mathbf{s_j},\mathbf{s_k}, \mathbf{s_l})=1+u(\mathbf{s_i}\cdot\mathbf{s_j}+\mathbf{s_j}\cdot\mathbf{s_k}+\mathbf{s_k}\cdot\mathbf{s_l}+\mathbf{s_l}\cdot\mathbf{s_i})+v(\mathbf{s_i}\cdot\mathbf{s_k}
 +\mathbf{s_j}\cdot\mathbf{s_l})+w\bigl[(\mathbf{s_i}\cdot\mathbf{s_j})(\mathbf{s_k}\cdot\mathbf{s_l})+(\mathbf{s_j}\cdot\mathbf{s_k})(\mathbf{s_l}  \cdot\mathbf{s_i})\bigr].
$$
Note that all possible pairings of the face spins appear except 
$(\mathbf{s_i}\cdot\mathbf{s_k})(\mathbf{s_j}\cdot\mathbf{s_l})$, which would allow for intersecting 
loops. The trace operator $\mathsf{Tr}$ in \eqref{eq:OnSquare} is defined again by 
\eqref{eq:Trace}. Note that the model is given here at the isotropic point, a simplification first 
given by \citet{Blote1989}. 

\begin{figure}[htb]
 \begin{center}
  \includegraphics[width=0.3\textwidth]{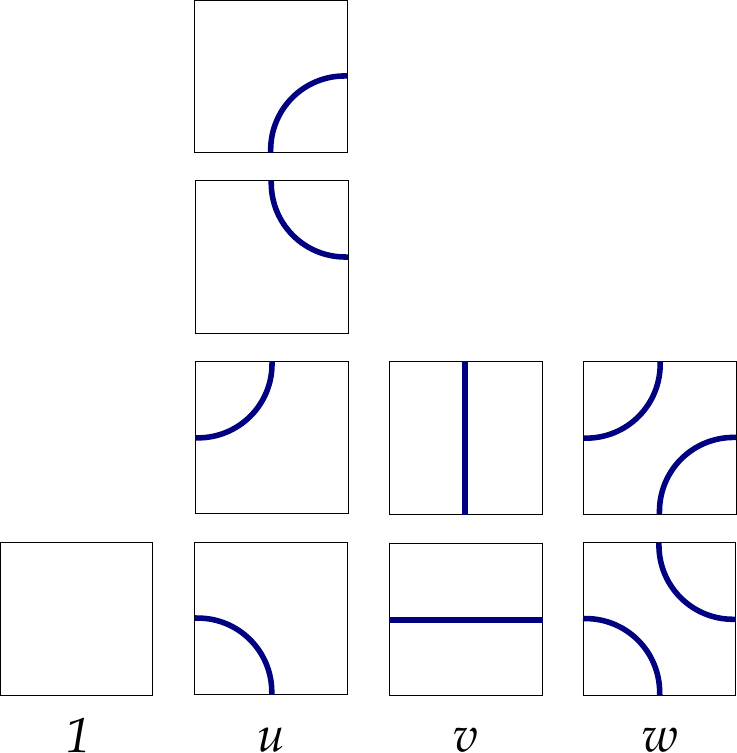}
  \caption[The Boltzmann weights.]{\label{fig:Weights}The Boltzmann weight of the nine possible 
  faces.}
 \end{center}
\end{figure}

Using the properties \eqref{eq:TrProp} of the trace, the partition function \eqref{eq:OnSquare} 
becomes
\begin{equation}
 \label{eq:OnSquareLoop}
 Z=\sum_\mathcal{L}u^{n_u}v^{n_v}w^{n_w}n^N.
\end{equation}
As before, $\mathcal{L}$ is the set of non-intersecting 
loop configurations and $n_u$, $n_v$ and $n_w$ are respectively the numbers 
of $u$, $v$ and $w$ type faces (see figure~\ref{fig:Weights}) in the configuration, and $N$ is still 
the total number of loops in $\mathcal{L}$. Again, $n$ is the loop fugacity 
while $u$, $v$ and $w$ are the weights of each face type.
In \citet{Blote1989}, the set of critical points is divided into five non-equivalent branches 
parametrized by $n$. We are interested in two of them here, corresponding 
respectively to the dilute and dense phases. They are both parametrized by
\begin{align}
 \label{eq:YBsol}
 u &= \frac{\sin2\lambda\sin\frac{3\lambda}{2}}{\sin2\lambda\sin3\lambda+\sin^2\frac{3\lambda}{2}}				\notag\\
 v &= \frac{\sin^2\frac{3\lambda}{2}}{\sin2\lambda\sin3\lambda+\sin^2\frac{3\lambda}{2}}					\\
 w &= \frac{\sin\frac{\lambda}{2}\sin\frac{3\lambda}{2}}{\sin2\lambda\sin3\lambda+\sin^2\frac{3\lambda}{2}}		\notag\\
 n &= -2\cos4\lambda,																	\notag
\end{align}
with the crossing parameter $\lambda$ constrained by
$0\leq\lambda\leq\frac{\pi}{4}$ or $\frac{\pi}{4}\leq\lambda\leq\frac{\pi}{2}$ for the dilute and 
dense phases, respectively. Note that, in both phases, the parameter 
$n$ covers the range $[-2,2]$ once. The weight of the empty face is normalized to $1$ for all 
$\lambda$. The original weights given in \citet{Blote1989} can be obtained by 
replacing $\lambda$ with $\frac{1}{2}(\pi-\theta)$ in \eqref{eq:YBsol}.

We conclude this subsection by recalling the
full weights of the dilute loop models \cite{Nienhuis1990} by including the spectral parameter
here indicated by $\hat{u}$ to distinguish it from the weight $u$ that appears in \eqref{eq:YBsol}
and figure~\ref{fig:Weights}.
The weight of the empty face is
\begin{alignat}{5}\label{eq:fullNien}
&&1+\frac{\sin\hat{u}\sin(3\lambda-\hat{u})}{\sin2\lambda\sin3\lambda}\notag\\
\intertext{while}
 &u_1=u_3=\displaystyle{\frac{\sin\hat{u}}{\sin3\lambda}},&\quad
 &u_2=u_4=\displaystyle{\frac{\sin(3\lambda-\hat{u})}{\sin3\lambda}}\notag
 \\[.3cm]
 &&v=\displaystyle{\frac{\sin\hat{u}\sin(3\lambda-\hat{u})}{\sin2\lambda\sin3\lambda}}&
 \\[.3cm]
 &w_1=-\displaystyle{\frac{\sin\hat{u}\sin(\lambda-\hat{u})}{\sin2\lambda\sin3\lambda}},&\quad
 &w_2=\displaystyle{\frac{\sin(2\lambda-\hat{u})\sin(3\lambda-\hat{u})}{\sin2\lambda\sin3\lambda}}\notag
\end{alignat}
where the weights of the faces in the second and fourth columns in 
figure~\ref{fig:Weights} are labeled from below by $u_1,u_2,u_3,u_4$ and $w_1,w_2$,
respectively. 
At the isotropic point $\hat{u}=3\lambda/2$, these weights reduce to the ones in (\ref{eq:YBsol}) 
after rescaling them to get the weight 1 for the empty face.

\subsection{Dilute logarithmic minimal models}\label{sec:DLM}

\begin{figure}[bht!]
 \begin{center}
  \subfigure[Dense phase configuration.]{\includegraphics[width=0.49\textwidth]{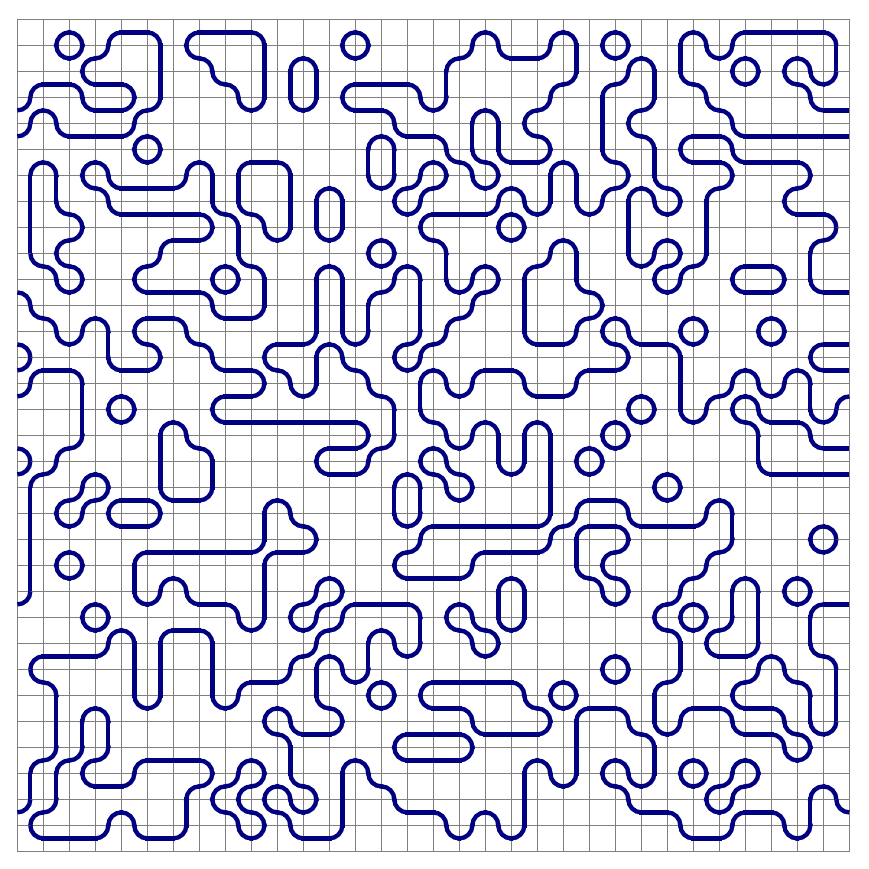}}\hfill
  \subfigure[Dilute phase configuration.]{\includegraphics[width=0.49\textwidth]{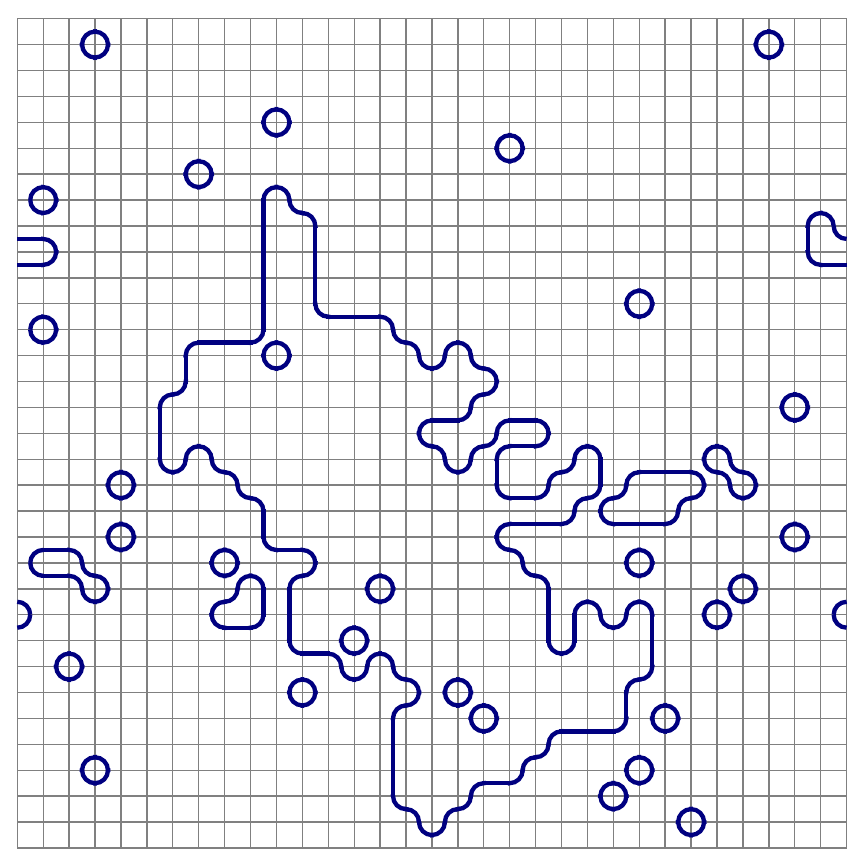}}
  \caption{\label{fig:denseVSdilute}Typical configurations on the $32\times32$ cylinder
  for the dense and dilute phases of 
  $\mathcal{DLM}(3,4)$ with $\beta_{\text{dense}}=\sqrt{2}$, $\beta_{\text{dilute}}=1$, and $c=1/2$ 
  for both. Note the large loop in the dilute configuration.
  In both phases, such large loops often occur.
}
 \end{center}
\end{figure}

The descriptions of physical systems through spin, Potts and loop models 
usually have different transfer matrices. 
One striking difference is the dimensions of the vector spaces upon 
which the spin or loop transfer matrices act; these dimensions are not equal. 
It is therefore not surprising that different continuum scaling limits may coexist for the same model, 
depending on the spin or loop description under study. By studying a dense 
loop representation of a family of models, different from the one used here, \citet{PRZ2006} 
found that some of the associated transfer matrices exhibit nontrivial Jordan blocks.
They subsequently argued that this gives rise to logarithmic CFTs in the
continuum scaling limit and labeled a 
two-parameter family of such limits by $\mathcal{LM}(p,p')$, for 
\emph{logarithmic minimal models}, where $p$ and $p'$ are positive coprime integers (see 
appendix~\ref{sec:LM}). Similarly, 
we believe that logaritmic CFTs arise in the continuum scaling limit of the loop models 
defined by \eqref{eq:OnSquareLoop}--\eqref{eq:fullNien} and thus label
the corresponding two-parameter family of continuum scaling limits 
as $\mathcal{DLM}(p,p')$, for \emph{dilute logarithmic minimal models}, where $p$ and $p'$ 
are as above. One way to justify the prefix 
``dilute'' is by the visual aspects of the loop configurations, as opposed to those 
of logarithmic minimal models in which only $w$-type faces are admissible. 
See figure~\ref{fig:denseVSdilute} for typical configurations of $\mathcal{DLM}(p,p')$ and 
appendix~\ref{sec:LM} for one of $\mathcal{LM}(p,p')$.

Because loops in both the dense and dilute phases of $\mathcal{DLM}(p,p')$ are likely to 
be related to conformal loop ensembles CLE$_\kappa$ (see 
\citet{Camia2006, Werner2005, Sheffield2006}), the relationship between the pair $(p,p')$, the 
parameter $\kappa$ (or $\bar{\kappa}=\kappa/4$) and the crossing parameter $\lambda$ 
needs to be given. The (logarithmic) CFT underlying $\mathcal{DLM}(p,p')$ has central charge
\begin{equation}
 \label{eq:centCharge}
 c=c(\bar{\kappa})=13-6\left(\bar{\kappa}+\frac{1}{\bar{\kappa}}\right),\qquad\text{where }\bar{\kappa}=\begin{cases} \frac{p}{p'}& \text{in the dilute phase} 
 \\[2mm] \frac{p'}{p}& \text{in the dense phase}\end{cases}
\end{equation}
and conformal weights
\begin{equation}
 \label{eq:confWeights}
 \Delta_{r,s}=\Delta_{r,s}(\bar{\kappa})=\frac{(\bar{\kappa}r-s)^2-(\bar{\kappa}-1)^2}{4\bar{\kappa}},\qquad r,s=1,2,3,\dotsc
\end{equation}
The usual duality $\kappa\leftrightarrow\frac{16}{\kappa}$ of CLE$_\kappa$ here becomes 
$\bar{\kappa}\leftrightarrow\frac{1}{\bar{\kappa}}$ and implies
\begin{equation}
 \label{eq:duality}
 c\left(\frac{1}{\bar{\kappa}}\right)=c(\bar{\kappa})\quad\text{and}\quad\Delta_{r,s}\left(\frac{1}{\bar{\kappa}}\right)=\Delta_{s,r}(\bar{\kappa}). 
\end{equation}
The link between $\mathcal{DLM}(p,p')$ and the $\mathcal{O}(n)$ model is completed by the expression of $\lambda$ in terms of $\bar{\kappa}$:
\begin{equation}
 \label{eq:lambBarKappa}
 \lambda(\bar{\kappa})=\frac{\pi}{2}\left(1-\frac{1}{2\bar{\kappa}}\right).
\end{equation}
This form for $\lambda(\bar{\kappa})$ is due to our choice of parametrization in \eqref{eq:YBsol};  one can also find 
$\lambda(\bar{\kappa})=\frac{\pi}{4\bar{\kappa}}$ in the literature.

In terms of $\bar{\kappa}$, the dilute and dense branches correspond to
\begin{equation}
 \label{eq:densDil}
 \begin{split}
  \frac{1}{2}\leq\bar{\kappa}\leq 1,\qquad	&	\text{dilute phase},	\\[1mm]
  1\leq\bar{\kappa}\leq\infty,\qquad		&	\text{dense phase}.
 \end{split}
\end{equation}
Moreover, we shall henceforth rename the loop gas fugacity $n$ by $\beta$:
\begin{equation}
 \label{eq:theBeta}
 \beta=-2\cos4\lambda=-2\cos\left(\frac{\pi}{\bar{\kappa}}\right).
\end{equation}

The duality $\bar{\kappa}\leftrightarrow\frac{1}{\bar{\kappa}}$ and its implication \eqref{eq:duality} 
suggest that the dense and dilute phases of $\mathcal{DLM}(p,p')$ are dual to each other. 
There also exists a link between $\mathcal{DLM}(p,p')$ and $\mathcal{LM}(p,p')$, which is 
twofold. To appreciate these dualities and links, we say that two models belong to the same 
universality class if they have the same central charge and share the same Kac table of conformal 
weights $\Delta_{r,s}(\bar{\kappa})$; this is typical of what is found in 
the CFT literature. (At this point, we do not require, or even address, that the models
are based on the same set of representations with identical Jordan-block structures.)
First, it is easy to see 
that the dense phase of $\mathcal{DLM}(p,p')$ lies in the same universality class as 
$\mathcal{LM}(p,p')$ since, for a given pair $(p,p')$, both models have the same 
$\bar{\kappa}$ (eq.~\eqref{eq:centCharge}), and thus the same central charge $c$ and Kac table. 
Second, the dilute phase of 
$\mathcal{DLM}(p,p')$ belongs to the same universality class as $\mathcal{LM}(p',p)$. This is also 
obvious since, by inverting $p\leftrightarrow p'$ in the definition of 
$\bar{\kappa}$ in the logarithmic minimal models, one falls back to the $\bar\kappa$ of the 
dilute phase of $\mathcal{DLM}(p,p')$. Consequently, the conformal weights 
and the central charges are also equal in these models. 

This equivalence has been known for a while: Duplantier \citep{Duplantier1987, Duplantier1989} 
used the Coulomb gas picture to discover a correspondence 
between the loop representation of the $\mathcal{O}(n)$ model and the Fortuin-Kasteleyn (FK) 
representation of the $n^2$-Potts model, which is valid for both the dilute and dense 
phases, when $n$ is in the physical regime, that is $n\in[0,2]$. Since $\mathcal{LM}$ is built upon 
the FK representation of the Potts model and $\mathcal{DLM}$ on the 
$\mathcal{O}(n)$ model, the equivalence merely appears by inheritance.

The principal series, i.e. models with $p'=p+1$, of the dilute logarithmic minimal models are
\begin{align}
 \label{eq:princSeries}
 \notag 	& \text{dense phase of }\mathcal{DLM}(p,p'):																															\\
 \notag 	& \textstyle\bar{\kappa}_{\mathrm{dense}}=\frac{p'}{p}=\left\{2,\frac{3}{2},\frac{4}{3},\frac{5}{4},\frac{6}{5},\dotsc,1\right\}\leftrightarrow\bigl\{\text{polymers, percolation, Ising, tricritical Ising, 3-Potts},\dotsc,\text{ 4-Potts}\bigr\}			\\
 		&																																						\\
 \notag 	& \text{dilute phase of }\mathcal{DLM}(p,p'):																															\\
 \notag	& \textstyle\bar{\kappa}_{\mathrm{dilute}}=\frac{p}{p'}=\left\{\frac{1}{2},\frac{2}{3},\frac{3}{4},\frac{4}{5},\frac{5}{6},\dotsc,1\right\}\leftrightarrow\bigl\{\text{polymers, percolation, Ising, tricritical Ising, 3-Potts},\dotsc,\text{ 4-Potts}\bigr\}.
\end{align}
In the case of $\mathcal{DLM}(1,1)$, $p$ and $p'$ are strictly speaking not coprime.
However, as indicated in \eqref{eq:princSeries}, the model can be viewed as arising
from the limit $p\to\infty$ of the principal series, for which
$p$ and $p'$ are coprime. For our purposes, though, it suffices to simply set $p=p'=1$.

The models in \eqref{eq:princSeries} are named according to their universality 
class. For instance, it is well known that the Ising model belongs to the universality 
class of $c=1/2$; this is why the models with 
$\bar{\kappa}_{\mathrm{dense}}=4/3$ ($\beta_{\text{dense}}=\sqrt{2}$) and 
$\bar{\kappa}_{\mathrm{dilute}}=3/4$ ($\beta_{\text{dilute}}=1$) are both called ``Ising 
model'' here. This might be confusing however since, in the $\mathcal{O}(n)$ model literature, 
one reserves that name for $\mathcal{O}(n=\beta=1)$, for both the dilute 
and dense phases. For comparison, a typical configuration for both phases of 
$\mathcal{DLM}(3,4)$ is shown in figure~\ref{fig:denseVSdilute}.

\begin{figure}[htb!]
 \begin{center}
  \includegraphics[width=0.5\textwidth]{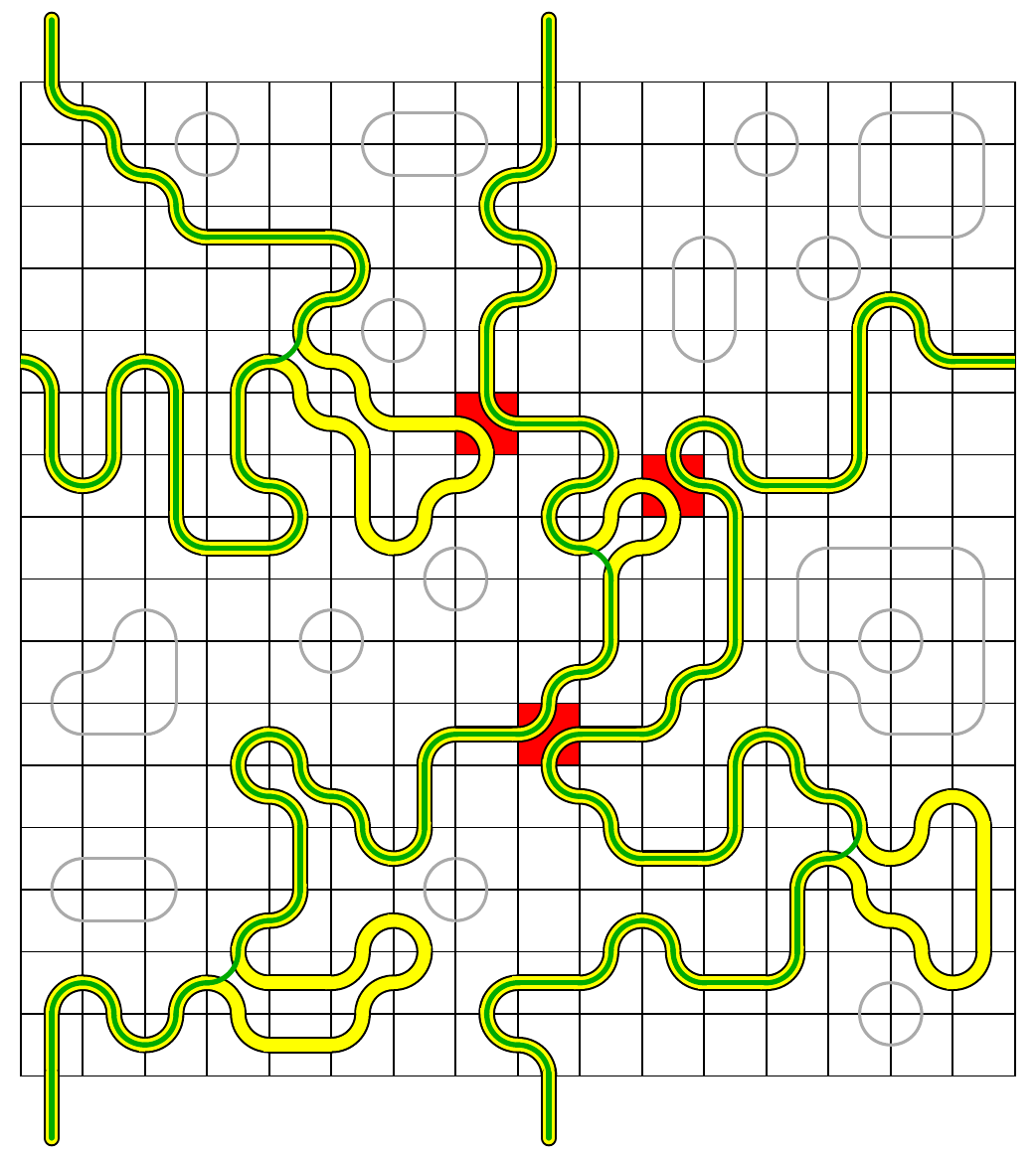}
  \caption[The observables.]{\label{fig:Observables}The three observables. This figure was 
  created specifically for the purpose of illustration and not randomly generated for a particular model.}
 \end{center}
\end{figure}

\subsection{Geometric fractal dimensions}\label{sec:geoFractalDim}

The \emph{hull perimeter}, or simply \emph{hull}, is defined as the set of all outer boundary sites of 
a cluster (see \citet{Grossman1986}). Since here there are no clusters, 
but only loops, this interpretation has to be adapted to the present models. To do this, we use the 
same trick as in \citep{Saint-Aubin2009} where it was applied to the $\mathcal{LM}$ models. 
We thus consider that the loops present in $\mathcal{DLM}$ configurations are 
Peierls contours of clusters of spins living on the lattice sites. 
Moreover we shall choose the boundary conditions such that at least one loop enters one of the boxes in the top row and exits in one of the bottom rows. This ``loop'' must then cross vertically the whole lattice and measurements can be done on this single object instead of taking averages over loops in each configuration. Such a loop will be called a {\em defect}. The analogy with the spin definition can then be constructed easily. The hull for a configuration with two defects is represented in pale (yellow) 
in figure~\ref{fig:Observables}.

The \emph{external perimeter} is the set of \emph{accessible} outer boundary sites of a cluster 
and constitutes a subset of the hull. In loop models, the external perimeter of 
a given loop can be interpreted as its contour minus the set of \emph{fjords}. For dilute models, a 
fjord is created each time the defect bounces on itself by using a $w$-type 
face, entrapping an area which now becomes inaccessible to the defect. (We have
borrowed the term {\em fjord} from \citet{Asikainen2003} where fjords with various gate sizes 
are considered. Ours have the narrowest gate size possible.) 
This is the discrete analogue of the situation 
where an SLE path engulfs a domain in its hull. 
It is emphasized that the definition of ``hull" used here is different from the one used in SLE.
The external perimeter is shown by a darker (green) line in figure~\ref{fig:Observables}, 
sitting above the paler hull.

The \emph{red bonds}, also known as \emph{singly connected bonds}, were first introduced by 
\citet{Stanley1977}. They are defined as the bonds of a cluster whose 
removal splits the cluster into two. If we imagine that an electrical current is flowing through the 
cluster, then the ``red bonds'' would be the first bonds to become red and melt, 
by analogy with a fuse. For loop gases on the cylinder,  
the previous definition needs to be adapted. Let us consider two clusters, percolating 
from top to bottom and delimited by two defects acting as Peierls contours, as in 
figure~\ref{fig:Observables}. (The cylindrical geometry is obtained by identifying the
left side with the right one.) Flipping one of the red colored $w$-type faces in 
figure~\ref{fig:Observables}
will create a unique ``cluster" encircling the cylinder. More generally, a \emph{red face} is 
any $w$-face which, by flipping it to its mirror-state, 
creates or destroys an encircling cluster. In other words, any $w$-type face formed by 
quarter-circles coming from {\em both} defects, is a red face. We will continue to 
refer to these $w$-faces as ``red bonds'' instead of ``red faces''. 

The conjectured fractal dimensions discussed in the following are taken from 
\citep{Duplantier2000}. Some of these expressions had been proposed before, in 
\citep{Saleur1987} and others, for the FK representation of Potts models. The hull fractal 
dimension is
\begin{equation}
 \label{eq:hull}
 d_h=1+\frac{\bar{\kappa}}{2}=2-2\Delta_{0,\pm1}
 =\begin{cases} 2-2\Delta_{p',p\pm1},\quad&\mathrm{dilute\ phase},
 \\[.2cm]
 2-2\Delta_{p,p'\pm1},\quad&\mathrm{dense\ phase},
 \end{cases}
\end{equation}
where $\Delta_{r,s}$ is defined as in \eqref{eq:confWeights} even for $r$ or $s$ non-positive.
\citet{Beffara2008} has shown rigorously that SLE$_\kappa$ paths have this dimension. The fact 
that the continuum scaling limit of the hull is actually an SLE$_\kappa$ path is known rigorously 
only for a handful of $\bar{\kappa}$ values. 

The dimension of the external perimeter is conjectured to be
\begin{equation}
 \label{eq:EP}
 d_{ep}=1+\frac{1}{2\bar{\kappa}}\theta(\bar{\kappa}-1)+\frac{\bar{\kappa}}{2}\theta(1-\bar{\kappa})
 =\begin{cases} 2-2\Delta_{p',p\pm1},\quad&\mathrm{dilute\ phase},
 \\[.2cm]
 2-2\Delta_{p\pm1,p'},\quad&\mathrm{dense\ phase},
 \end{cases}
\end{equation}
where $\theta$ is the Heaviside step function for which $\theta(0)=\tfrac{1}{2}$. 
In the dilute phase $\frac{1}{2}\leq\bar{\kappa}\leq 1$, $d_{ep}=d_h=1+\frac{\bar{\kappa}}{2}$, 
which can be understood in 
terms of CLE$_\kappa$. Indeed, the dilute interval is $2\leq\kappa\leq 4$ and, 
for these values of $\kappa$, it is known that the contours (loops) are 
almost surely simple, implying that there is almost surely no intersection of the contour with 
itself. This was shown rigorously by \citet{Rohde2005} (see also \citet{Kager2004}).

Finally, the fractal dimension of the \emph{red bonds} is conjectured to be 
\begin{equation}
 \label{eq:redBonds}
d_{rb}=1+\frac{\bar{\kappa}}{2}-\frac{3}{2\bar{\kappa}}
 =2-2\Delta_{0,\pm2}
 =\begin{cases} 2-2\Delta_{p',p\pm2},\quad&\mathrm{dilute\ phase},
 \\[.2cm]
 2-2\Delta_{p,p'\pm2},\quad&\mathrm{dense\ phase}.
 \end{cases}
\end{equation}
This fractal dimension is negative on the dilute interval. At first, one might think that 
\eqref{eq:redBonds} is amiss, or that it does not hold for all 
$\frac{1}{2}\leq\bar{\kappa}\leq1$. But as explained in \citet{Mandelbrot1990}, negative fractal 
dimensions may occur in certain physical problems. We will discuss further 
the issue with negative fractal dimensions in section~\ref{sec:redBonds}.

We gather in table \ref{tab:paramsModels} these predicted values for a selection of dilute models. 
One of the main goals of this paper is to verify some of these predictions.
\begin{table}[htb!]
 \begin{center}
  \begin{tabular}{|c||c|c|c||c|c|c|}
   \hline 						&				&							&										&						&										\\[-2mm]
   model (dilute phase)				&	$\bar{\kappa}$	&	$\beta$					&	$c$									&	$d_h=d_{ep}$			&	$d_{rb}$								\\[2mm]
   \hline\hline 					&				&							&										&						&										\\[-2mm]
   name, $\mathcal{DLM}(p,p')$		&	$\frac{p}{p'}$	&	$-2\cos\frac{\pi}{\bar{\kappa}}$	&	$13-6\left(\bar{\kappa}+\frac{1}{\bar{\kappa}}\right)$	&	$1+\frac{\bar{\kappa}}{2}$	&	$1+\frac{\bar{\kappa}}{2}-\frac{3}{2\bar{\kappa}}$	\\[2mm]
   \hline 						&				&							&										&						&										\\[-2mm]
   percolation, $\mathcal{DLM}(2,3)$		&	$\frac{2}{3}$	&	$0$						&	$0$									&	$\frac{4}{3}$			&	$-\frac{11}{12}$							\\[2mm]
   Ising, $\mathcal{DLM}(3,4)$			&	$\frac{3}{4}$	& 	$1$						&	$\frac{1}{2}$							&	$\frac{11}{8}$			&	$-\frac{5}{8}$							\\[2mm]
   tricritical Ising, $\mathcal{DLM}(4,5)$	&	$\frac{4}{5}$	&	$\sqrt{2}$					&	$\frac{7}{10}$							&	$\frac{7}{5}$			&	$-\frac{19}{40}$							\\[2mm]
   3-Potts, $\mathcal{DLM}(5,6)$		&	$\frac{5}{6}$	&	$\frac{1}{2}\bigl(\sqrt{5}+1\bigr)$	&	$\frac{4}{5}$							&	$\frac{17}{12}$			&	$-\frac{23}{60}$							\\[2mm]
   4-Potts, $\mathcal{DLM}(1,1)$		&	$1$			&	$2$						&	$1$									&	$\frac{3}{2}$			&	$0$									\\[1mm]
   $\mathcal{DLM}(5,7)$				&	$\frac{5}{7}$	&	$\frac{1}{2}\bigl(\sqrt{5}-1\bigr)$	&	$\frac{11}{35}$							&	$\frac{19}{14}$			&	$-\frac{26}{35}$							\\[2mm]
   \hline 						&				&							&										&						&										\\[-2mm]
   $\mathcal{DLM}(3,5)$				&	$\frac{3}{5}$	&	$-1$						&	-$\frac{3}{5}$							&	$\frac{13}{10}$			&	$-\frac{6}{5}$							\\[2mm]
   polymers, $\mathcal{DLM}(1,2)$		&	$\frac{1}{2}$	&	$-2$						&	$-2$									&	$\frac{5}{4}$			&	$-\frac{7}{4}$							\\[1mm] \hline
  \end{tabular}
 \end{center}
 \caption{\label{tab:paramsModels}Parameters and fractal dimensions for different dilute models. 
 The models with $\beta<0$ were not simulated in this work but added here for comparison.}
\end{table}


\section{Measurements of fractal dimensions of dilute models}\label{sec:fractalDimension}

\subsection{The Minkowski fractal dimension of the defect}\label{sec:Minkowski}

Let $S$ be a subset of $\mathds{R}^d$ and let $\epsilon$ be the mesh of a hypercubic lattice 
drawn on $\mathds{R}^d$. The Minkowski or box-counting definition of the 
fractal dimension of $S$ is given by
\begin{equation}
 \label{eq:DfMink}
 d_S=\lim_{\epsilon\to 0}\frac{\ln{N(\epsilon)}}{\ln{1/\epsilon}},
\end{equation}
where $N(\epsilon)$ is the number of boxes that intersect $S$. In the present case, $d=2$ and 
$\mathds{R}^2$ is replaced by a bounded subset of area 
$A$. It is natural to introduce a function $R(H,V)$ that measures the linear size of $A$ in terms of 
the mesh $\epsilon$. Here $H$ and $V$ are the numbers of horizontal and vertical boxes in the lattice. The simplest definition of $R$ is given implicitly by
\begin{equation}
 \label{eq:Rnaif}
 A=\bigl(R(H,V)\epsilon\bigr)^2.
\end{equation}
The Minkowski fractal dimension is then $d_S=\displaystyle\lim_{H,V\to\infty}d_S^{H\times V}$ 
where
\begin{equation}
 \label{eq:Dfasym}
 d_S^{H\times V}=\ln{N(H,V)}\Bigl/\bigl(\ln{R(H,V)}-\frac{1}{2}\ln A\bigr)\Bigr..
\end{equation}
Of course, the numbers $H$ and $V$, the area $A$ and the mesh $\epsilon$ 
are related. For simplicity, we will use $A=1$.

The definition \eqref{eq:Rnaif} of the linear size $R(H,V)$ is natural for studying the defect of 
$\mathcal{LM}(p,p')$ (see appendix~\ref{sec:LM}) as in 
\citep{Saint-Aubin2009}. Indeed, \emph{(i)} each state of a box contains precisely two 
quarter-circles and \emph{(ii)} each box of the lattice of area $A$ is accessible to the 
defect. Neither of these two observations holds for the dilute models $\mathcal{DLM}(p,p')$. 
Beside the states containing two quarter-circles ($w$-faces), there are some 
that contain a line segment of length $\epsilon$ ($v$-faces) or only one quarter-circle ($u$-faces) or 
nothing at all (the empty face). We shall use boundary conditions on the subset $S$ 
that forbid loops to reach the boundary. In this case, not all states are available for the 
boundary boxes. There are therefore two problems to resolve. 

The first problem is to decide the number of boxes of side $\epsilon$ needed to cover each 
box state. Should the $u$- and $v$-faces occupy the same area? What about 
the $w$-faces? The second problem is to make the definition of the area $A$ precise.
One could simply decide 
that each box of the lattice may contain at most two quarter-circles so that 
$A=2HV\epsilon^2$, where $\epsilon$ is the side length of a box small enough to cover one 
quarter-circle. But one might want to reduce this number of boxes due to the fact 
that $w$-faces are not allowed along the boundary. Or one can decide 
to define the area as that corresponding to the number 
$N_{\text{max}}(\epsilon)$ of $\epsilon$-boxes necessary to cover the largest subset of the lattice 
that can be occupied by the defect. The length of the longest defect, that can be 
drawn on the lattice, times $\epsilon^2$ would be the total area $A$ that any 
observable could possibly occupy, leading to a maximal fractal dimension of $2$. Note 
that the definition of the area using $N_{\text{max}}$ depends on the answer given to the first 
question. In what follows, we choose this interpretation and explore various 
definitions of $R$ in the case of the hull.

It is hoped that all reasonable choices will lead to the same fractal dimension for the defects in 
the limit $H,V\rightarrow\infty$. Still, some choices might be more 
appropriate for finite lattices and the rest of this section is devoted to see the impact of various 
choices on the quality of the extrapolation. Note that the following discussion is for a strip with the boundary conditions just described.
However, it can easily be adapted to lattices with cylindrical geometry.

\begin{figure}[tbh]
 \begin{center}
  \subfigure[]{\label{fig:DfCorresp1}\includegraphics[width=0.20\textwidth]{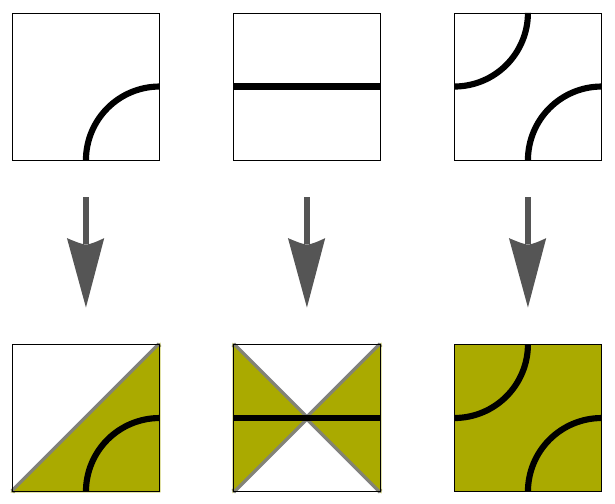}}\hspace{1.5cm}
  \subfigure[]{\label{fig:Res2A}\includegraphics[width=0.3\textwidth]{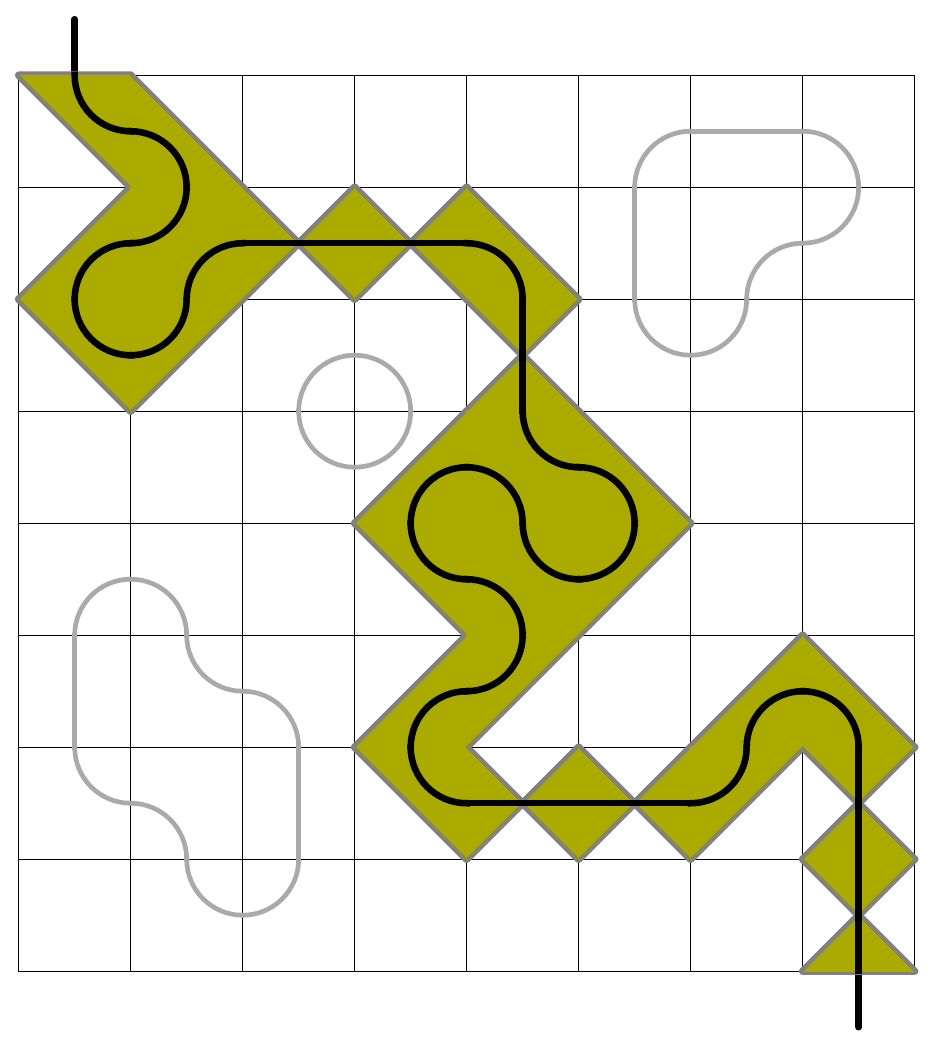}}
  \caption{\label{fig:Corr1}Covering of the defect for $R=R_1^{\left(\frac12,\frac12\right)}$ of a sample $8\times 8$ lattice.}
 \end{center}
\end{figure}
\begin{figure}[tbh]
 \begin{center}
  \subfigure[]{\label{fig:DfCorresp2}\includegraphics[width=0.20\textwidth]{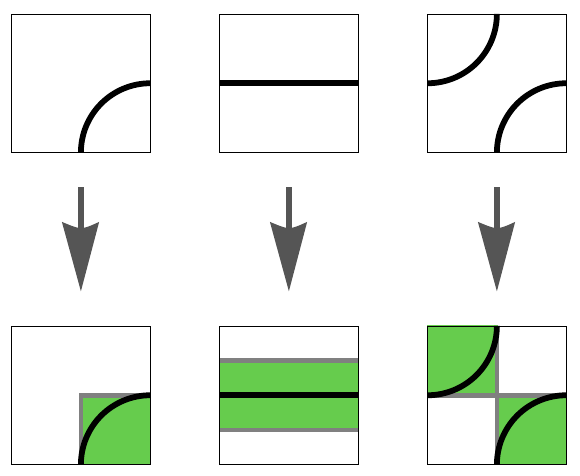}}\hspace{1.5cm}
  \subfigure[]{\label{fig:Res1B}\includegraphics[width=0.3\textwidth]{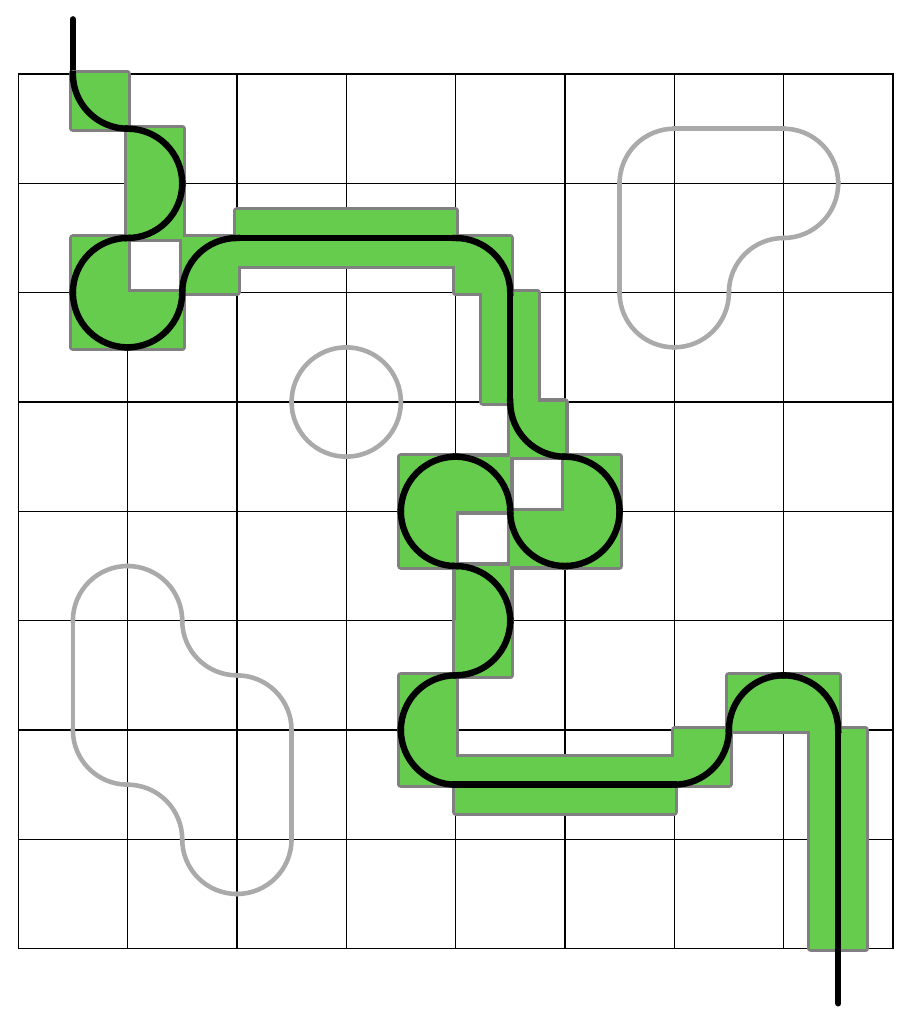}}
  \caption{\label{fig:Corr2}Covering of the defect for $R=R_2^{\left(\frac14,\frac12\right)}$ of a sample $8\times 8$ lattice.}
 \end{center}
\end{figure}

As a first example of the various weights given to boxes crossed by the defect, consider boxes with 
only one quarter-circle or a straight line segment as half-filled, and boxes with 
two quarter-circles as completely filled. More precisely, if $\epsilon^2$ is the area of a tile, then 
$s_u=s_v=\frac{1}{2}\epsilon^2$ and $s_w=\epsilon^2$, where $s_i$ is the 
area occupied by an $i$-face. A graphical interpretation of this choice is 
given in figure~\ref{fig:DfCorresp1}. If the numbers $H$ and $V$ are even, 
the defect with maximal $N_{\text{max}}(\epsilon)$ covers the lattice as follows: all boxes not 
touching the boundary are $w$-faces, while all boundary boxes but one are $u$- or $v$-faces. 
The last boundary box is necessarily the empty face. We conclude that, for this choice, the linear 
size $R$ is given by
\begin{equation}
 \label{eq:2A}
 R=\left(HV-H-V+3/2\right)^{1/2}.
\end{equation}
More generally, one may define the two-parameter family 
\begin{equation}
 \label{eq:2Afamily}
 R_1^{\left(a,b\right)}=\bigl(2a(HV-H-V+1)+b\bigr)^{1/2},
\end{equation}
valid when $H$ and $V$ are even, such that $s_u=a\epsilon^2$, $s_v=b\epsilon^2$ and 
$s_w=2a\epsilon^2$, with $0<a\leq b<2a$. Then $R$ in \eqref{eq:2A} is simply 
$R_1^{\left(\frac12,\frac12\right)}$. 
Figure~\ref{fig:Res2A} shows the ``area'' covered by a defect 
using this correspondence.
Note that, unless $a=\frac12$, $\epsilon^2$ is not in general the area of a box.

One might also be interested in cases where $b\geq2a>0$. Now the $v$-faces cover a larger area 
than the $w$-faces, and defects with $N_{\text{\rm max}}$ are found in configurations different 
from those leading to $R_1$. We thus define
\begin{equation}
 \label{eq:1Bfamily}
 R_2^{\left(a,b\right)}=\bigl(2aV+b(HV-H-2V+3)\bigr)^{1/2}.
\end{equation}
This family of coverings contains the choice $s_u=\frac14\epsilon^2$, 
$s_v=s_w=\frac12\epsilon^2$, or $R_2^{\left(\frac14,\frac12\right)}$, depicted in 
figure~\ref{fig:Corr2}. We found this covering to be another relevant choice since it can be 
argued that a quarter-circle is shorter than a straight line segment, and thus that $s_u$ should be smaller than $s_v$. Because the range of validity 
of \eqref{eq:2Afamily} and \eqref{eq:1Bfamily} do not overlap, we might as 
well define
\begin{equation}
 \label{eq:theR}
 R_{(a,b)}=\begin{cases}R_1^{\left(a,b\right)}, & 0<a\leq b<2a \\[3mm] 
 R_2^{\left(a,b\right)},\ & b\geq2a>0\end{cases}
\end{equation}
which is valid as above only when $H$ and $V$ are even numbers, 
and for the strip only. 
Although we limited our experiments in what follows to members of 
\eqref{eq:theR}, one could in principle extend this definition to include the less intuitive situations 
where $0<b<a$ or $s_w\neq2s_u$.

\begin{figure}[htb]
 \begin{center}
  \includegraphics[width=0.8\textwidth]{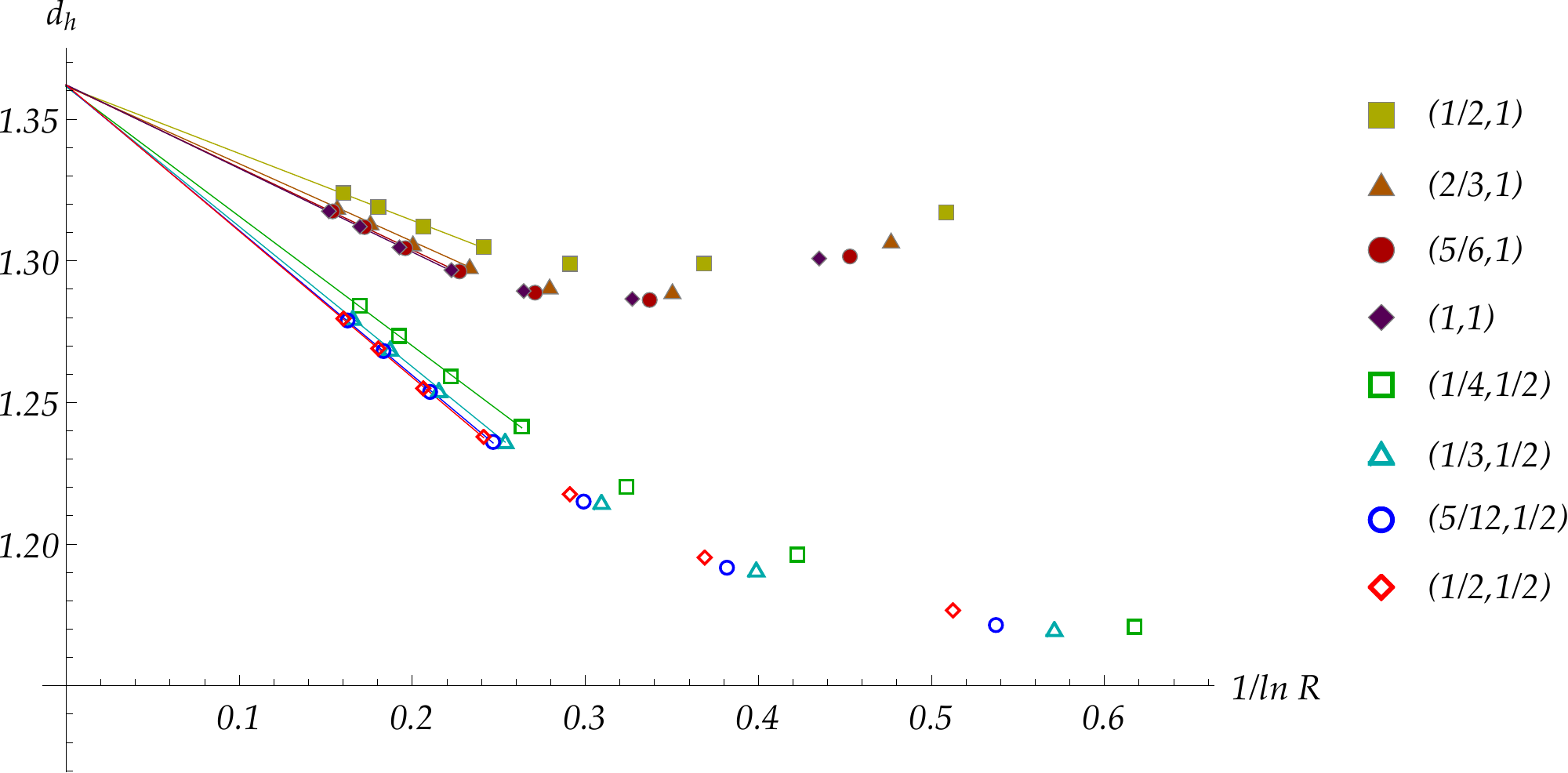}
  \caption{\label{fig:defDf}Linear fits of the hull dimension $\widehat{d_h}^{H\times V}$ against 
  $1/\ln R$ for different covering definitions using the model $\mathcal{DLM}(3,4)$ on the square 
  geometry. The pair $(a,b)$ refers to $R_{(a,b)}$.}
 \end{center}
\end{figure}
Figure~\ref{fig:defDf} shows fits of the same data obtained from various definitions of $R$. To do 
these, the numbers of $u$, $v$ and $w$-faces were stored separately for every 
configuration measured. We see that all the linear fits of $d_h$ extrapolate to 
about $1.362$ at $1/\ln R=0$. The exact value is of no importance right 
now since these measurements are imprecise and the lattices used are relatively small. It seems 
clear, however, that $b$ is the important parameter and that, when $b=1$, the 
$\widehat{d_h}^{H\times V}$ values
are closer to their asymptotic values. For example, the two curves 
with $R_{\left(\frac12,1\right)}$ and $R_{\left(\frac12,\frac12\right)}$ are 
far apart, but all those with $b=1$ (or $b=1/2$) are bunched together: this probably means that 
the $v$-faces play a significant role in the length of the defect. Because the results
are closer to their asymptotic values for $b=1$, 
we choose to work with the covering $(a,b)=(1,1)$.


\subsection{Technical issues}

\subsubsection{Simulations on the cylinder}\label{sec:Convergence}

As explained before, the convergence to the asymptotic value of the fractal dimension of the defect 
is very slow. Despite the improvements made to the algorithm (cf. 
section~\ref{sec:Improvements}), lattices of linear size larger than $H=512$ are difficult to reach. 
Let us explore the cylindrical geometry using the model $\mathcal{DLM}(3,4)$, 
where $\beta=1$. Since no loop counting is required in this case, the algorithm is simpler and thus 
performs more Monte Carlo cycles per second.
\begin{figure}[htb]
 \begin{center}
  \includegraphics[width=0.8\textwidth]{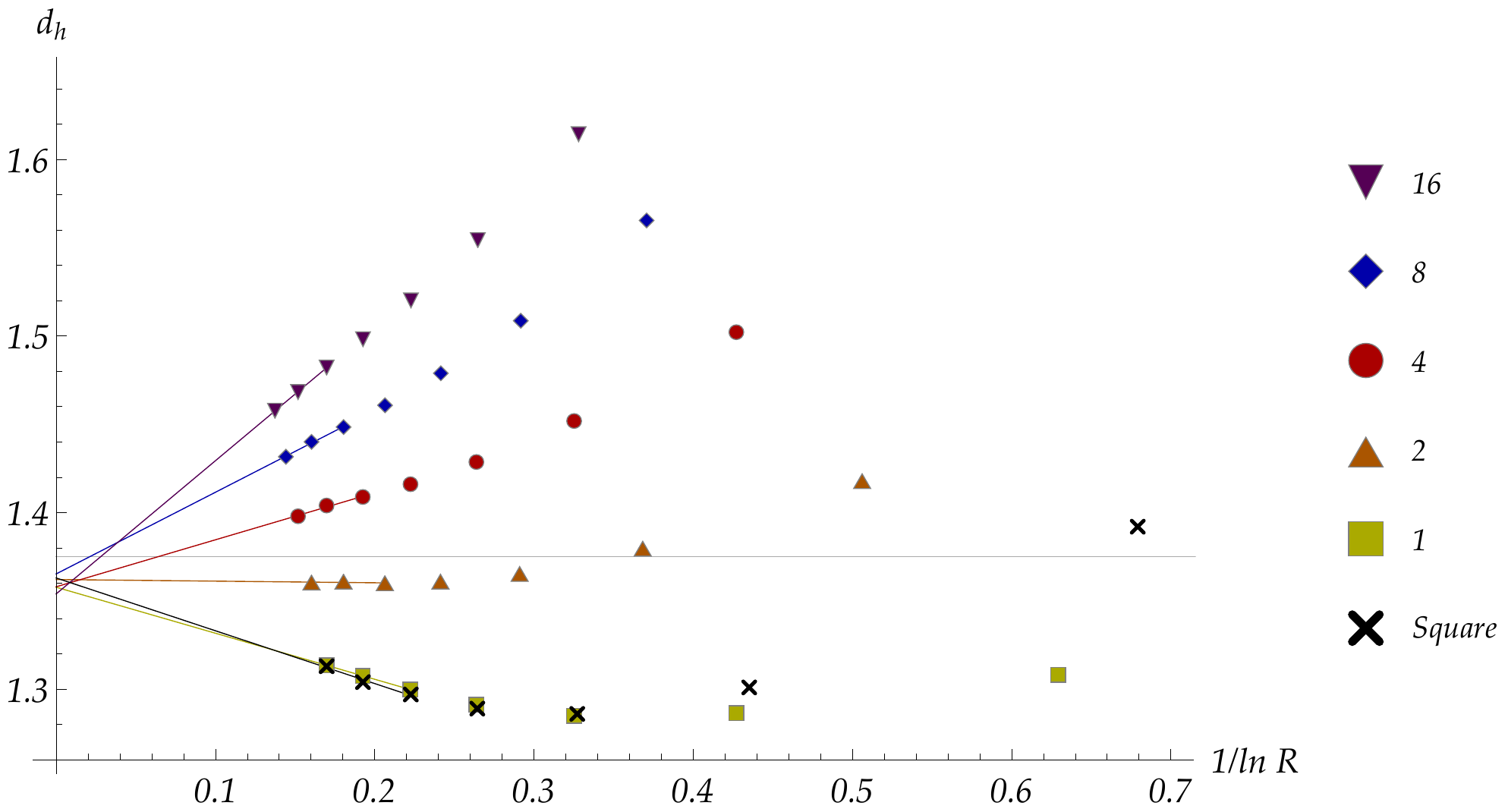}
  \caption{\label{fig:DflnR}Linear fits of $\widehat{d_h}^{H\times V}$ as a function of $1/\ln R$ for 
  cylinders of different $V/H$ ratio and for the square. The horizontal line marks $d_h=11/8$ for
  $\mathcal{DLM}(3,4)$.}
 \end{center}
\end{figure}
We measured the fractal dimension of the defect on cylinders with different aspect ratios $V/H\in\{1,2,4,8,16\}$, where $V$ is the number of boxes along the symmetry axis 
of the cylinder. The entry and exit points of the defect were set in columns corresponding to the 
azimuthal angle being $0$ and $\pi$. For each of these ratios $V/H$, the hull fractal dimensions were obtained for $H\in\{4,8,16,32,64,128,256\}$. The plot of linear fits with respect to $1/\ln R$ is 
shown in figure~\ref{fig:DflnR}. Obviously, these fits are coarse. Nevertheless, it is already clear that they 
converge to a narrow window around the predicted value of equation 
\eqref{eq:hull}, $d_h=\frac{11}8=1.375$ (horizontal line), as does the curve corresponding to the 
measurements on the square strip. Still, none seems to aim right at $1.375$. 
Furthermore, the cylindrical geometry might not speed up the convergence towards the 
asymptotic value $d_h$. 

These results call for some discussion. First, the results for the two geometries, that of the strip and of the cylinder, behave quite differently. On the 
longest cylinder $V/H=16$, the $\widehat{d_h}^{H\times V}$'s are significantly higher 
than $d_h=1.375$, as opposed to those of the square geometry which are all lower, except for $H=4$. Second, 
boundary effects cannot be the only reason for the difference between the two 
geometries. As discussed in section~\ref{sec:BoundEffects}, significant changes in $d_h$ 
are seen close to the boundary only in a region whose width is a fraction of 
$H$. Therefore, the long cylinders $V/H=4, 8$ and $16$ should have approximately the same 
$\widehat{d_h}^{H\times V}$'s. What we see is probably the effect of 
configurations with larger and larger winding number. The defect can indeed wrap around the 
cylinder as it progresses along it. Intuitively, one would think that defects winding, 
say, twice around the cylinder are longer in average than those with no winding. Because the 
number of configurations with large winding numbers increases with the ratio 
$V/H$, a longer cylinder will have defects with larger $\widehat{d_h}^{H\times V}$. Third, one 
might be tempted to do the measurements of $d_h$ on cylinders with 
$V/H=2$, because the measurements for coarse meshes are slightly closer to the predicted value 
and the slope for larger meshes is small. Of course, if the choice for these 
cylinders was based on this reason, this would be cheating and it is not clear that the fit over the 
whole range of $R$ would be any better since it would also require non-linear terms.

\subsubsection{Boundary effects and the advantage of the cylindrical geometry}\label{sec:BoundEffects}

We now explore the effects of the boundary on the measurements of observables. As the lines 
below will show, it is possible that, in a certain limit, $d_h$ might actually be a 
function of the distance from the boundary. We shall try to give a proper definition of the fractal 
dimension as a function of this distance, but let us first consider the effect of 
joining two lattices with different fractal dimensions. Suppose that two square lattices of the 
same linear size $H$ are covered by different models, $\mathcal{DLM}(p,p')$ 
and $\mathcal{DLM}(q,q')$ such that $\frac{p}{p'}\neq\frac{q}{q'}$, whose defects have dimension 
$d_{h1}$ and $d_{h2}$, respectively, with $d_{h1}>d_{h2}$. And consider
configurations on a new lattice, obtained by joining the two above. Which $\beta$ should give the weight of the
loops that overlap the two sublattices? This is a subtle problem and, for the 
sake of the present argument, it might be enough to confine loops to either sublattices. For 
sufficiently large $V$ and $H$, the part of the defect in the first half will be of length 
$c_1H^{d_{h1}}$ and $c_2H^{d_{h2}}$ in its second half, for some constants $c_1$ and $c_2$. 
Thus, its total length is 
$c_1H^{d_{h1}}\left(1+H^{(d_{h2}-d_{h1})}c_2/c_1\right)$ and the fractal dimension in the whole 
region is
\begin{equation}
 \label{eq:dH1dH2}
 \left.\ln\Bigl[c_1H^{d_{h1}}\bigl(1+H^{(d_{h2}-d_{h1})}c_2/c_1\bigr)\Bigr]\right/\ln H\underset{H\rightarrow\infty}{\longrightarrow} d_{h1},
\end{equation}
where $R$ is approximated by $H$. Note that this argument can easily be modified to describe 
two lattices of unequal sizes. This shows a simple fact about fractal dimensions: if a geometric 
object $S$ is studied on a region $D\subset\mathds{R}^d$ of a finite lattice, then its fractal 
dimension 
$d_S^D$ over $D$ will be equal to $d_S^{E_\text{max}}$, that is to the fractal dimension over the 
subregion $E_\text{max}\subset D$ where it is maximal.

The present situation is slightly more complicated as $d_h$ might be varying continuously with the distance from the boundary. What does this mean? The definition \eqref{eq:DfMink} 
applies to a subset $S\subset\mathds{R}^2$ or to a subset $S$ of the cylinder as a whole. To 
define a ``local'' fractal dimension, consider a cylinder of fixed ratio 
$r=\text{length}/\text{perimeter}$. Let $L$ be the length of the cylinder and $0\le l_1<l_2\le L$ and 
let $C_{12}$ be the annulus along (or band around) the cylinder containing points 
at a distance $l$ from one extremity with $l_1<l<l_2$. Then one can measure 
$d_h$ on $C_{12}$ by replacing $\mathds{R}^d$ with $C_{12}$ in the discussion leading to 
\eqref{eq:DfMink}. Numerical estimates of $d_h$ can be obtained by covering the cylinder with larger and 
larger $H\times V$ lattices, with $r=V/H$, and counting the number of intersections of the defect in 
a given $C_{12}$. Note that the defect can 
meander out of a given annulus, come back to it, leave again, and so on; all of its intersections 
with $C_{12}$ must be counted. Unfortunately this fractal dimension 
$d_h=d_h(l_1,l_2)$ is difficult to measure numerically. The narrowest annulus on the small 
cylinder $H\times V=8\times 8$ already takes up one eighth of the cylinder's length. 
Since the fractal dimension is obtained as an extrapolation over several lattice sizes, the smallest 
section of the cylinder where this $d_h(l_1,l_2)$ can be measured effectively 
is of length $l_2-l_1=L/8=(V\epsilon)/8$. Unfortunately, some quick explorations have shown to us 
that a circle at a distance $L/8$ from the boundary is almost free of the boundary effects. 
The following experiment allows to probe boundary effects.

\begin{figure}[tbh]
 \begin{center}
  \includegraphics[width=0.9\textwidth]{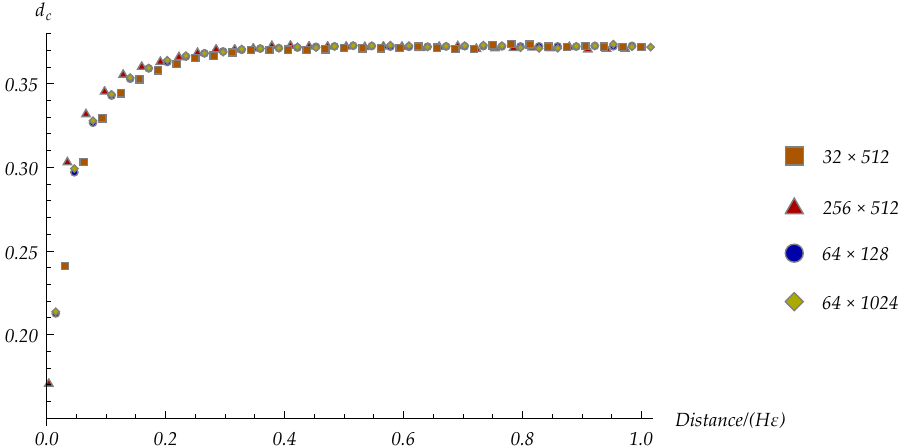}
  \caption{\label{fig:DfalgCyl} Fractal dimension of the intersections between the defect and circles 
  $C$ on cylinders \mbox{$(H,V)=(32,512), (256,512), (64,128)$ and $(64,1024)$} for 
  $\mathcal{DLM}(3,4)$.}
 \end{center}
\end{figure}

Let $C$ be a circle at a fixed distance from the boundary of the cylinder. This circle $C$ has 
(standard or fractal) dimension one. 
We consider the set $c\subset C$ of intersections of the defect with $C$. The fractal dimension 
$d_c$ of this set is therefore $0\le d_c\le d_C=1$. It is this definition that is 
used in figure~\ref{fig:DfalgCyl} where $d_c$ is measured as a function of the distance from the 
nearest extremity, in units of 
$H\epsilon$. Cylinders with $(H,V)=(32,512), (256,512), (64,128)$ and $(64,1024)$ were used. 
All the rows of the cylinders were measured, but only those at a distance 
$0\leq l\leq H\epsilon$ are shown. Moreover, only a subset of the available data is drawn; the 
size of the datasets for $64\times128$ and $64\times1024$ was reduced by a 
factor of $2$ while the one for $256\times512$ was by $8$. Two observations clearly stand out 
from figure~\ref{fig:DfalgCyl}: the fractal dimension seems to ``feel'' the 
boundary roughly up to a distance of $l/(H\epsilon)=\frac13$ and this distance is independent of 
both the ratio $V/H$ of the cylinder and its mesh size $\epsilon$. 
The following simple argument relates $d_c$ to the fractal dimension $d_h$ of the hull.

Let $N(H,V)$ be the number of intersections of the defect with $H$ rows chosen far away from the 
boundaries of a long 
cylinder. (We use the notation of section~\ref{sec:Minkowski}.) Choose the circle $C$ to be one of 
these $H$ rows. On average, there will be $N/H$ intersections with this circle 
$C$ and the fractal dimension $d_c$ introduced above is the limit of 
$$
 \frac{\ln (N(H,V)/H)}{\ln R_C(H)}
$$ 
as $H\rightarrow\infty$. As before, $R_C(H)$ stands for the number of ``boxes'' necessary to 
cover $C$. We choose this number to be the maximal number of intersections 
with $C$, namely $2H-1$. However, $d_h$ is the limit of $\ln N(H,V)/\ln R_H(H,V)$ where now 
$R_H(H,V)$ is the square root of the maximal number of intersections of the defect with the $H$ 
rows, that is approximately $\sqrt{2H^2}$. So, for large $H$'s, $d_h\sim\ln N/\ln H$ and
$$
 d_c=\frac{\ln (N(H,V)/H)}{\ln (2H-1)}\sim\frac{\ln N}{\ln H}-1=d_h-1=\frac{\bar{\kappa}}{2}.
$$
Since $d_h\in [1,2]$, the dimension $d_c$ lies in $[0,1]$ as desired and, for 
$\mathcal{DLM}(3,4)$, $d_c$ should be $\frac38$, very close to the value $0.37$ seen in 
figure~\ref{fig:DfalgCyl}.

\begin{figure}[h!]
 \begin{center}
  \includegraphics[width=0.9\textwidth]{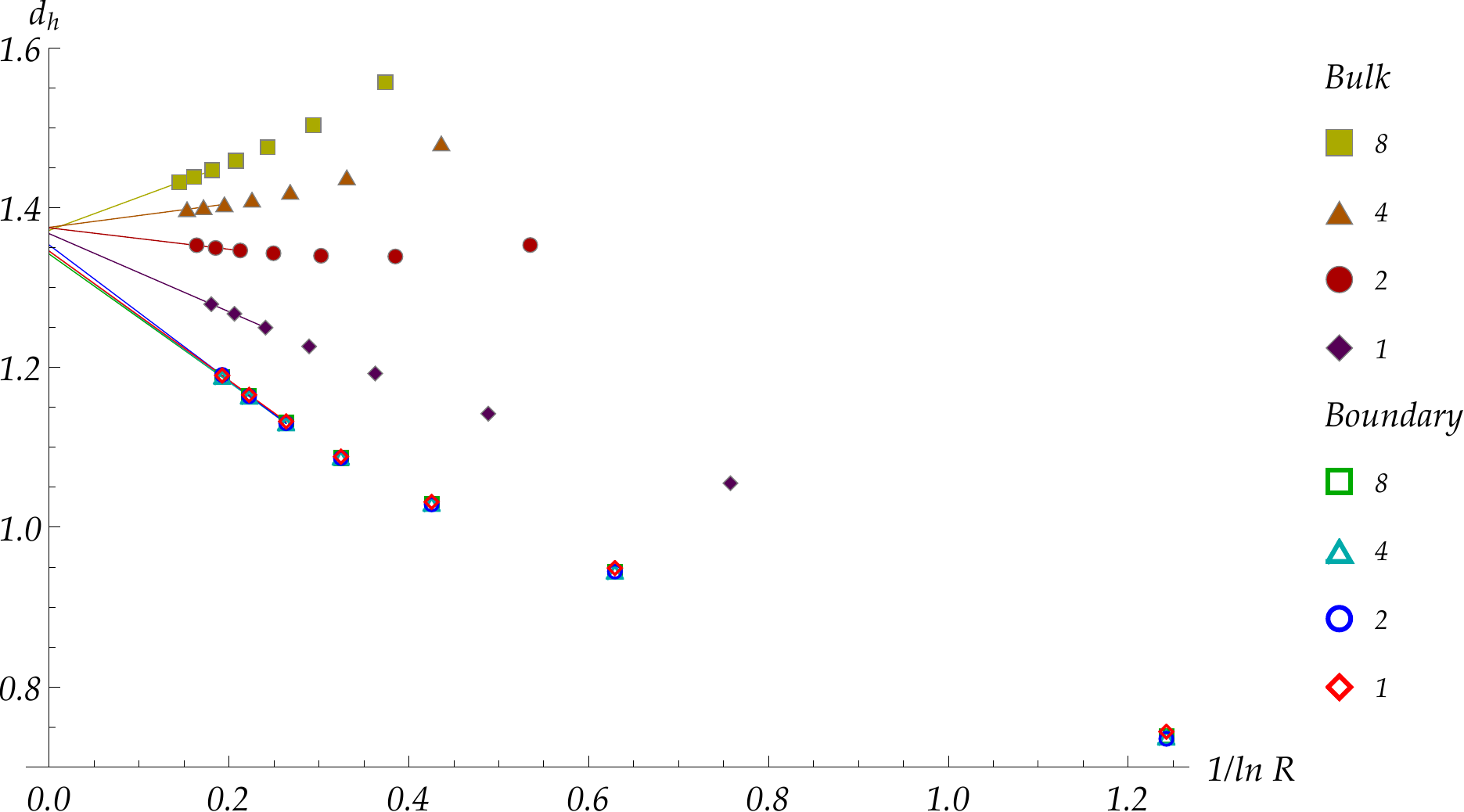}
  \caption{\label{fig:frontBulk}Measurements of $C_{\text{boundary}}$ and $C_{\text{bulk}}$ on 
  cylinders with $V/H=1,2,4$ and $8$.}
 \end{center}
\end{figure}

This result suggests yet another experiment; one in which $d_h$ is measured on the annulus 
$C_{\text{bulk}}$ which is defined for $l_1/L=H/(3V)$ and $l_2/L=1-l_1/L$, that is, an annulus 
excluding the region where 
boundary effects are felt. Figure~\ref{fig:frontBulk} shows the data for the measurements of $d_h$ 
on $C_{\text{bulk}}$ for cylinders with ratios $V/H\in\{1,2,4,8\}$. There are four other sets of points 
all converging to the same value, around $1.35$. These 
are the measured $\widehat{d_h}$ on the complement $C_{\text{boundary}}$ of $C_{\text{bulk}}$ 
in the cylinder. Results remain preliminary as they are limited to 
$H\le 256$ only. But they are striking. The finite fractal dimensions on $C_{\text{bulk}}$ all 
converge quite close to the expected $\frac{11}8$ and all those on the 
complement to another lower value. These results do not allow us to determine whether the fractal 
dimension on $C_{\text{boundary}}$ and that on $C_{\text{bulk}}$ are 
distinct. Indeed, in figure~\ref{fig:DfalgCyl}, the points for the largest lattice 
($256\times 512$) clearly stand over those of the other lattices when the circle $C$ is 
chosen close to the boundary. They seem to indicate that the limit of $d_c$ is not yet reached. But 
even if these two fractal dimensions on $C_{\text{boundary}}$ and 
$C_{\text{bulk}}$ were the same, these results do allow to conclude that the rate of convergence to 
their asymptotic value is different. This is the main conclusion of this long analysis.

Figure \ref{fig:frontBulk} shows how to measure efficiently any $d_S$ in the bulk whenever the rate of 
convergence of $d_S^{H\times V}(l)$ depends on the distance $l$ from the 
boundary. If we assume that the function $d_S^{H\times V}(l)$ is approximately conformally 
invariant, then the effects of the boundary will be definitely more important for measurements 
on the square. To see this, map conformally the cylinder onto the disk by the exponential. Because 
the circumference of the cylinder has $H$ sites, the annulus that is at 
distance $H/3$ from the extremity will be sent onto a circle with radius 
$r\sim e^{-2\pi/3}\,r_{\text{disk}}<0.125 \, r_{\text{disk}}$. The Schwarz-Christoffel map that sends 
the disk onto the square will change slightly the form of this inner circle close to the center, but it 
will not change the fact that a minute number of boxes of the square 
geometry actually lie in the bulk (see \citet{Langlands2000}), approximately one hundredth of the 
total number of boxes. This precludes obtaining a good measurement of 
$d_S$ in the bulk using square geometries. Indeed, even for a $256\times256$ lattice, one would 
have to count the intersections of the defect with a small square at the 
center of the lattice of about $30\times 30$ boxes.

We therefore propose to restrict our study to the fractal dimension $d_S$ in the bulk, and use the 
cylindrical geometry to measure it, with cylinders of ratio $V/H=2$. For 
simplicity and to minimize boundary effects in our experiments, we define the bulk 
as the region enclosed by the annuli at distance $l_1=H/2$ and $l_2=3H/2$ from one 
of the extremities of the cylinder. 
To extend the discussion in section \ref{sec:Minkowski}, we must seek the maximum number $N_{\text{\rm max}}$ of intersections the defect can have with this bulk section. Again we assume that faces $u$ and $v$ count for one intersection and $w$ for two. For $(a,b)=(1,1)$, the most dense configuration has $H(2H-1)$ such intersections and we define $R$ to be the square root of this number.
This choice of the size ratio $V/H=2$ and bulk region allows 
to use half of the boxes for measurements with only a reasonable increase in the total number of 
boxes. This seems to be a good compromise. 

\subsubsection{The distribution of the winding number}\label{sec:windingNumber}

\begin{figure}[bht!]
 \begin{center}
  \subfigure[Winding number $\alpha=0$.]{\includegraphics[width=0.35\textwidth]{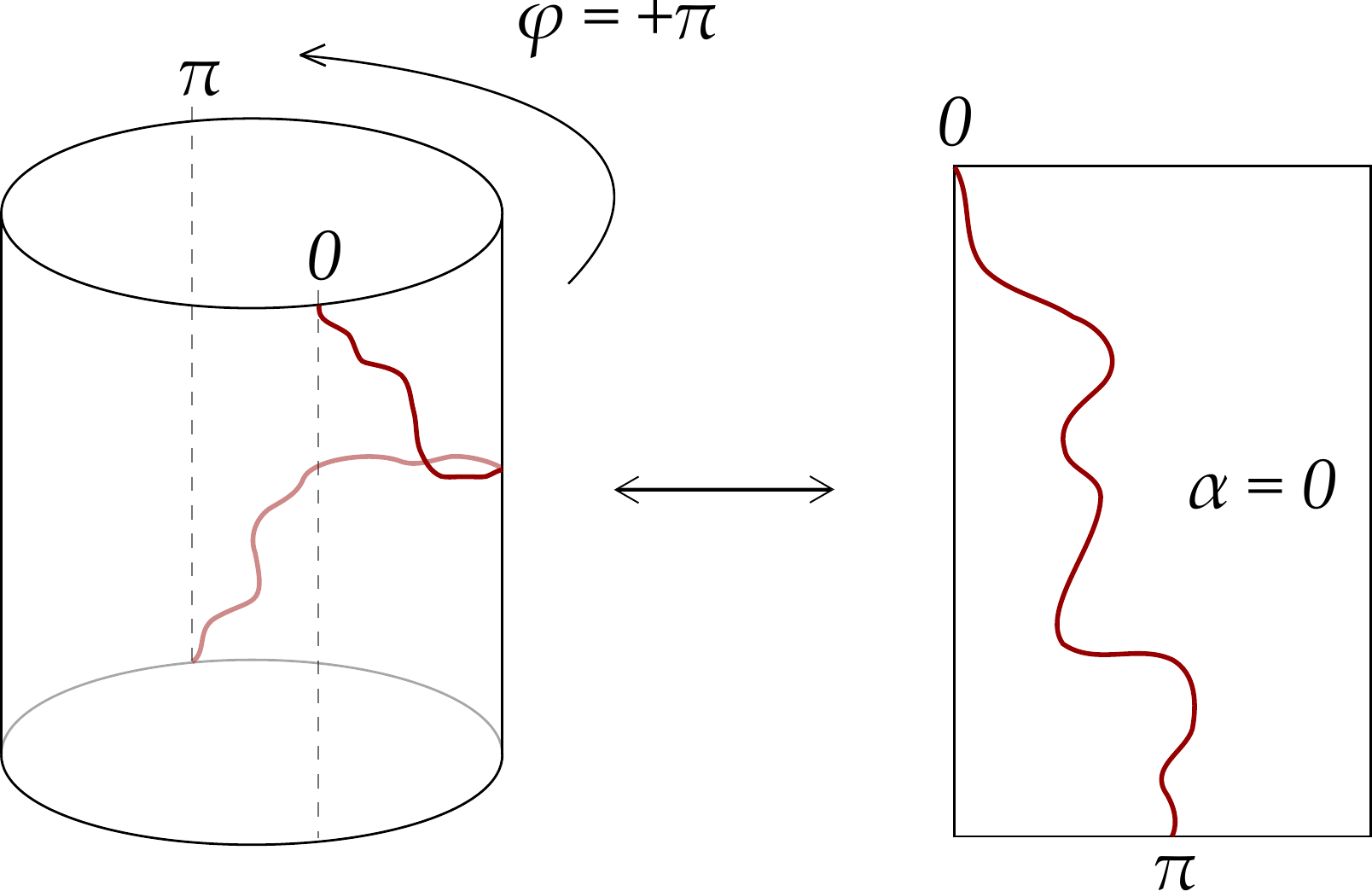}}\hspace{0.1\textwidth}
  \subfigure[Winding number $\alpha=1$.]{\includegraphics[width=0.49\textwidth]{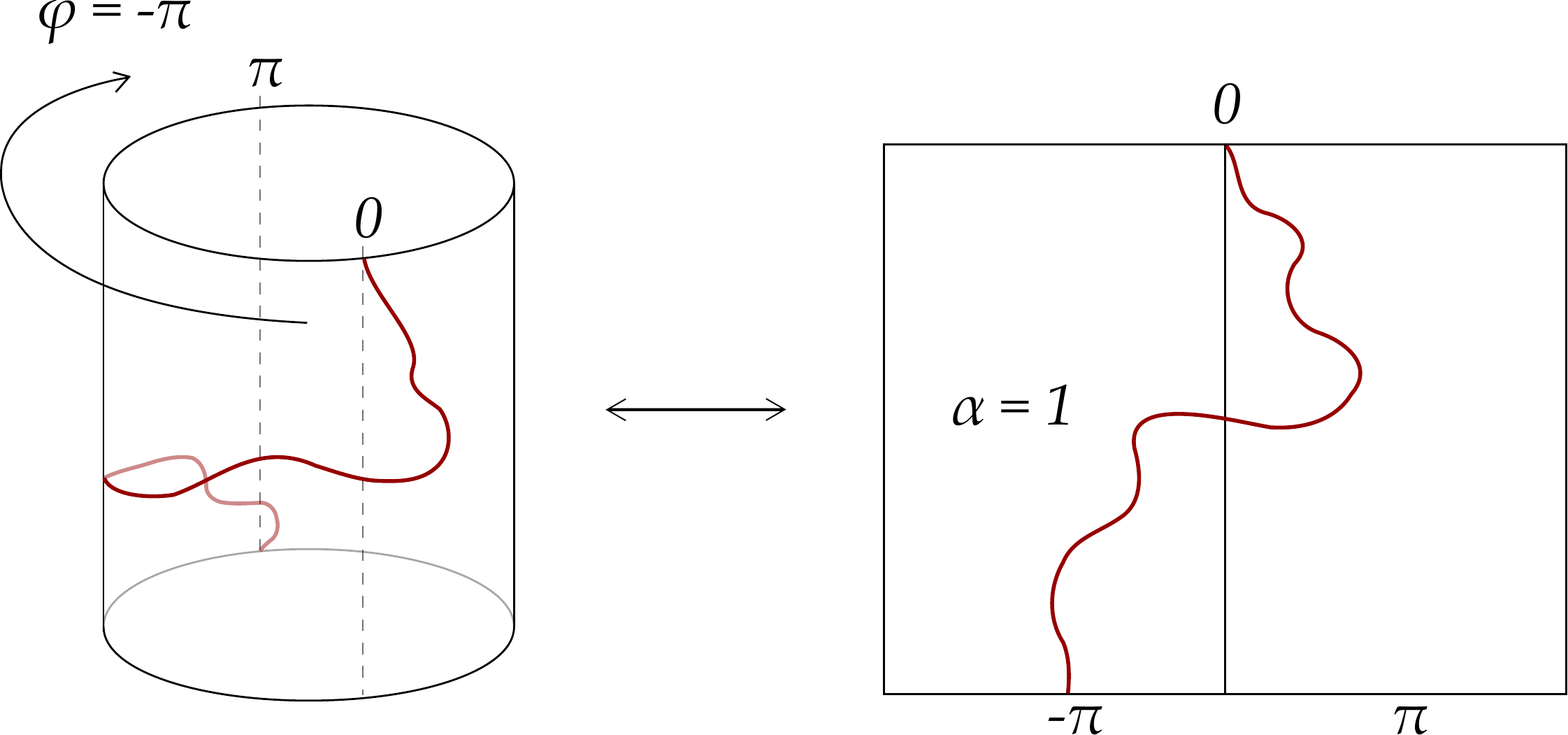}}
  \caption{\label{fig:alpha01}Sample configurations with winding numbers $\alpha=0$ and $\alpha=1$.}
 \end{center}
\end{figure}

\begin{figure}[h!]
 \begin{center}
  \includegraphics[width=0.6\textwidth]{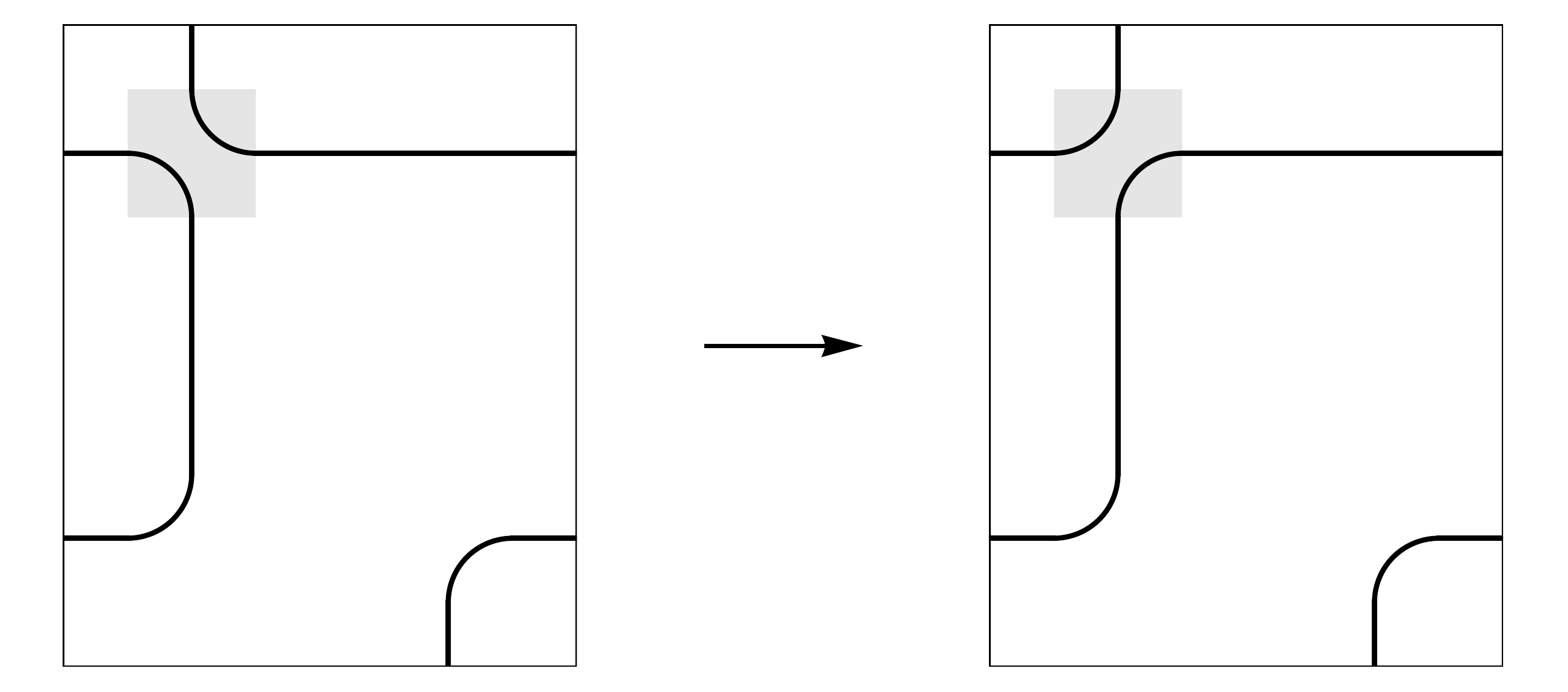}
  \caption{\label{fig:winding}The algorithm preserves the parity of the winding number.}
 \end{center}
\end{figure}

Recall that the entry and exit points of the defect are chosen to be at locations corresponding to 
azimuthal angles equals to $0$ and $\pi$, respectively. We define the \emph{winding number} 
$\alpha$ of the defect by
\begin{equation}
 \label{eq:windingNumber}
 \alpha(\varphi) = \frac{\pi-\varphi}{2\pi},
\end{equation}
where $\varphi$ is the azimuthal angle of the defect. A configuration will be said to have winding 
number $\alpha=0$ if the angle of the defect $\varphi$ has increased by 
$\pi$ while going from one extremity to the other. Note that, with this definition, a variation of $-\pi$ 
of the angle would amount to a winding number $\alpha=1$. Two sample 
configurations having winding numbers $0$ and $1$ are shown in figure~\ref{fig:alpha01}. The 
definition \eqref{eq:windingNumber} is asymmetric, and rather unpleasant. 
However, as will be seen soon, this asymmetry is actually welcome. The simulations reported 
in the previous paragraphs \ref{sec:Convergence} and \ref{sec:BoundEffects} 
were done starting the thermalization with a defect with zero winding number. 
This has the consequence that no configurations with odd 
winding numbers are considered in the Monte Carlo integration. Indeed, the upgrade algorithm 
(section~\ref{sec:Algorithm}) always changes the winding number by zero or two units. 
For example, in figure~\ref{fig:winding}, the flip of the gray box changes the winding number from 
$0$ to $2$. 

The exclusion of all configurations with an odd winding number is a serious problem. One could 
try to overcome it by measuring separately the distributions of the defect's 
length for configurations with even and for those with odd winding numbers. For the latter, it would 
amount to start with a configuration in this set. But then the determination 
of the distribution of the defect length for the whole space of configurations would raise the 
question of the relative probability of the two sets and this seems extremely 
difficult. As it turns out, this problem does not occur! Indeed, the probability distribution of the 
winding number is obviously invariant through a mirror containing the cylinder 
axis, corresponding to $\varphi\rightarrow -\varphi$. Due to our definition, however, this symmetry
sends a configuration with winding number $0$ onto a configuration with winding number $1$. 
More generally, configurations with winding number $\alpha$ are sent onto configurations with 
winding number $1-\alpha$ and are 
therefore equiprobable. Note that, if the entry and exit points had been put 
in the same column, that is with the same azimuthal angle, the problem of discarding the odd or 
the even configurations would have been a serious one.

\begin{figure}[tbh]
 \begin{center}
  \includegraphics[width=0.6\textwidth]{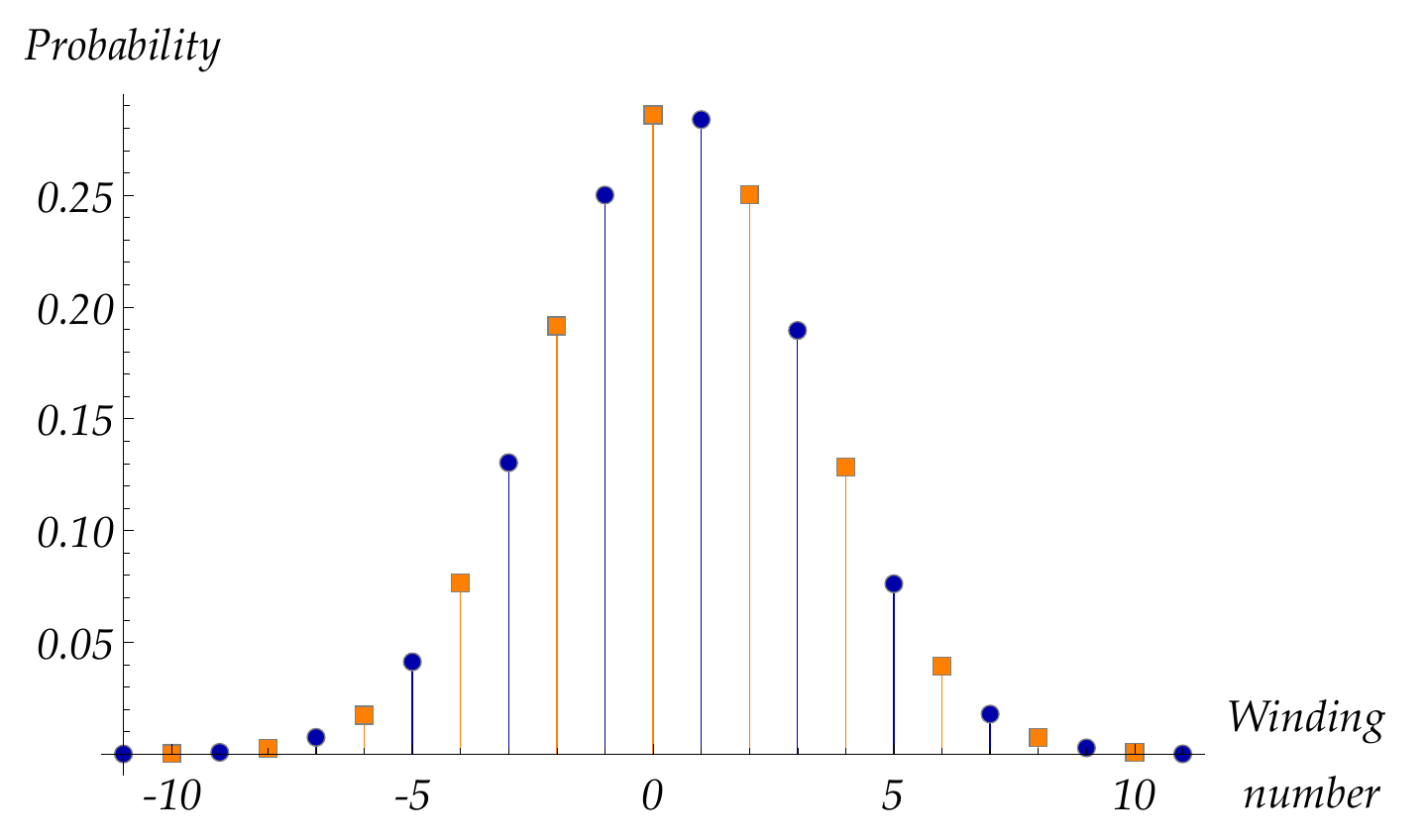}
  \caption{\label{fig:windingDistribution}The discrete probability density of the winding number on 
  the $V/H=16$ cylinder for $\mathcal{DLM}(3,4)$.}
 \end{center}
\end{figure}

Figure~\ref{fig:windingDistribution} shows the probability density function of each winding number 
within its set, either even (square dots) or odd (disks), measured on the 
cylinder with ratio $V/H=16$ and $H=8$ for the model $\mathcal{DLM}(3,4)$. It supports quite 
nicely the previous observation, as the two histograms are mapped onto one 
another by the symmetry $\alpha\leftrightarrow 1-\alpha$. It is interesting to note also that quite large 
winding numbers are reached. The histogram box for $10$ is readable, but 
larger winding numbers were also reached. \citet{Pinson1994} and \citet{Arguin2002} have 
considered the probability of Fortuin-Kasteleyn configurations having a given 
homotopy on the torus. Their results may be applicable to long cylinders or, at least, give a 
reasonable prediction for this geometry.


\section{Results}\label{sec:Results}

To estimate the fractal dimension of the hull for a given lattice size, we took $m$ measurements, 
usually with $m=10\,000$, of the length $L$ of the defect in the bulk region 
of the $V/H=2$ cylinder. Each pair of consecutive measurements is separated by a fixed number $\Delta$ of 
Monte-Carlo iterations which varies according to the size of the lattice and 
the model itself. This procedure was carried out simultaneously on $n$ machines, with $n\geq20$, 
and for cylinders with $H=8,16,32,64,128,256,512$. The average fractal 
dimension of a given cylinder of size $H$ is then obtained by averaging the defect's length 
$N(H,V)=L$ of equation \eqref{eq:Dfasym} over all the $m\times n$ data:
\begin{equation}
 \label{eq:avValue}
 \widehat{d_h}=\frac{1}{mn}\sum_{j=1}^{n}\sum_{i=1}^{m}\frac{\ln L_{i,j}}{\ln R(H)},
\end{equation}
where $R(H)=\bigl(2H^2-H\bigr)^{1/2}$ in the bulk for the cylinders with $V/H=2$. The same 
procedure was used for the external perimeter. A typical dataset obtained by this 
method is shown in table \ref{tab:Meas34}.
\begin{table}[thb!]
 \begin{center}
  \begin{tabular}{|c||c|c|c|c|c|c|c|}
   \hline 				&					&					&					&					&					&					&					\\[-2mm]
   $H$					&	$8$				&	$16$				&	$32$				&	$64$				&	$128$				&	$256$				&	$512$				\\[2mm]
   \hline 				&					&					&					&					&					&					&					\\[-2mm]
   $\widehat{d_h}$		&	$1.2772\bigl|\bigr.6$	&	$1.2923\bigl|\bigr.4$	&	$1.3045\bigl|\bigr.4$	&	$1.3143\bigl|\bigr.4$	&	$1.3215\bigl|\bigr.5$	&	$1.3279\bigl|\bigr.7$	&	$1.3323\bigl|\bigr.4$	\\[2mm]\hline
  \end{tabular}
 \end{center}
 \caption{\label{tab:Meas34}Hull fractal dimension measurements for $\mathcal{DLM}(3,4)$ on 
 cylinders \mbox{$H={8,16,32,64,128,256,512}$} with $V/H=2$. The notation $1.2772\bigl|\bigr.6$ 
 means that the $95\%$ confidence interval's half width is $0.0006$ 
and centered on $1.2772$.}
\end{table}

\begin{table}[htb!]
 \begin{center}
  \begin{tabular}{|c||c|c|c||c|c|}
   \hline 						&				&							&						&					&					\\[-2mm]
   model name, $\mathcal{DLM}(p,p')$	&	$\bar{\kappa}$	&	$\beta$					&	$d_h=d_{ep}$			&	$\widehat{d_h}$		&	$\widehat{d_{ep}}$	\\[2mm]
   \hline\hline 					&				&							&						&					&					\\[-2mm]
   percolation, $\mathcal{DLM}(2,3)$		&	$\frac{2}{3}$	&	$0$						&	$\frac{4}{3}\simeq1.3333$	&	$1.3328\bigl|\bigr.13$	&	$1.3332\bigl|\bigr.19$	\\[2mm]
   Ising, $\mathcal{DLM}(3,4)$			&	$\frac{3}{4}$	& 	$1$						&	$\frac{11}{8}=1.3750$		&	$1.3757\bigl|\bigr.13$	&	$1.3751\bigl|\bigr.15$	\\[2mm]
   tricritical Ising, $\mathcal{DLM}(4,5)$	&	$\frac{4}{5}$	&	$\sqrt{2}$					&	$\frac{7}{5}=1.4000$		&	$1.4025\bigl|\bigr.14$	&	$1.3986\bigl|\bigr.19$	\\[2mm]
   3-Potts, $\mathcal{DLM}(5,6)$		&	$\frac{5}{6}$	&	$\frac{1}{2}\bigl(1+\sqrt{5}\bigr)$	&	$\frac{17}{12}\simeq1.4167$	&	$1.4197\bigl|\bigr.11$	&	$1.4119\bigl|\bigr.12$	\\[2mm]
   4-Potts, $\mathcal{DLM}(1,1)$		&	$1$			&	$2$						&	$\frac{3}{2}=1.5000$		&	$1.5021\bigl|\bigr.12$	&	$1.4615\bigl|\bigr.10$	\\[2mm]
   $\mathcal{DLM}(5,7)$				&	$\frac{5}{7}$	&	$\frac{1}{2}\bigl(\sqrt{5}-1\bigr)$	&	$\frac{19}{14}\simeq1.3571$	&	$1.3592\bigl|\bigr.20$	&	$1.3586\bigl|\bigr.26$	\\[2mm]\hline
  \end{tabular}
 \end{center}
 \caption{\label{tab:Meas}Extrapolated fractal dimensions for the hull and external perimeter. Fits of the form $\beta_0+
 \beta_1/\ln R+\beta_2/\ln^2 R+\beta_3/\ln^3 R$ were considered and model testing led us to set $\beta_2=0$ (see section \ref{sec:modelTesting}).}
\end{table}
The extrapolated fractal dimensions of the hull and external perimeter for all models studied are 
shown in table \ref{tab:Meas} with their 95\% confidence interval.  The 
meaning of our confidence interval is as follows: if the experiment was repeated many times, then 
an average of $5$ fits out of $100$ would predict an average value 
$\widehat{d}$ at $1/\ln R\to0$ that lies outside the $95\%$ confidence interval of table 
\ref{tab:Meas}. We see that the extrapolations are in good agreement with the 
theoretical values given by equations \eqref{eq:hull} and \eqref{eq:EP}. Still, some of the 
confidence intervals do not contain the theoretical values; this is especially true for 
the external perimeter in $\mathcal{DLM}(1,1)$. Even though the definition of the confidence 
interval does not imply that it should contain the theoretical value with $95\%$ 
probability, one would like to better understand these small departures. These discrepancies for the hull and
the external perimeter will be discussed in the next subsections. 

\begin{figure}[htb!]
 \begin{center}
  \includegraphics[width=0.9\textwidth]{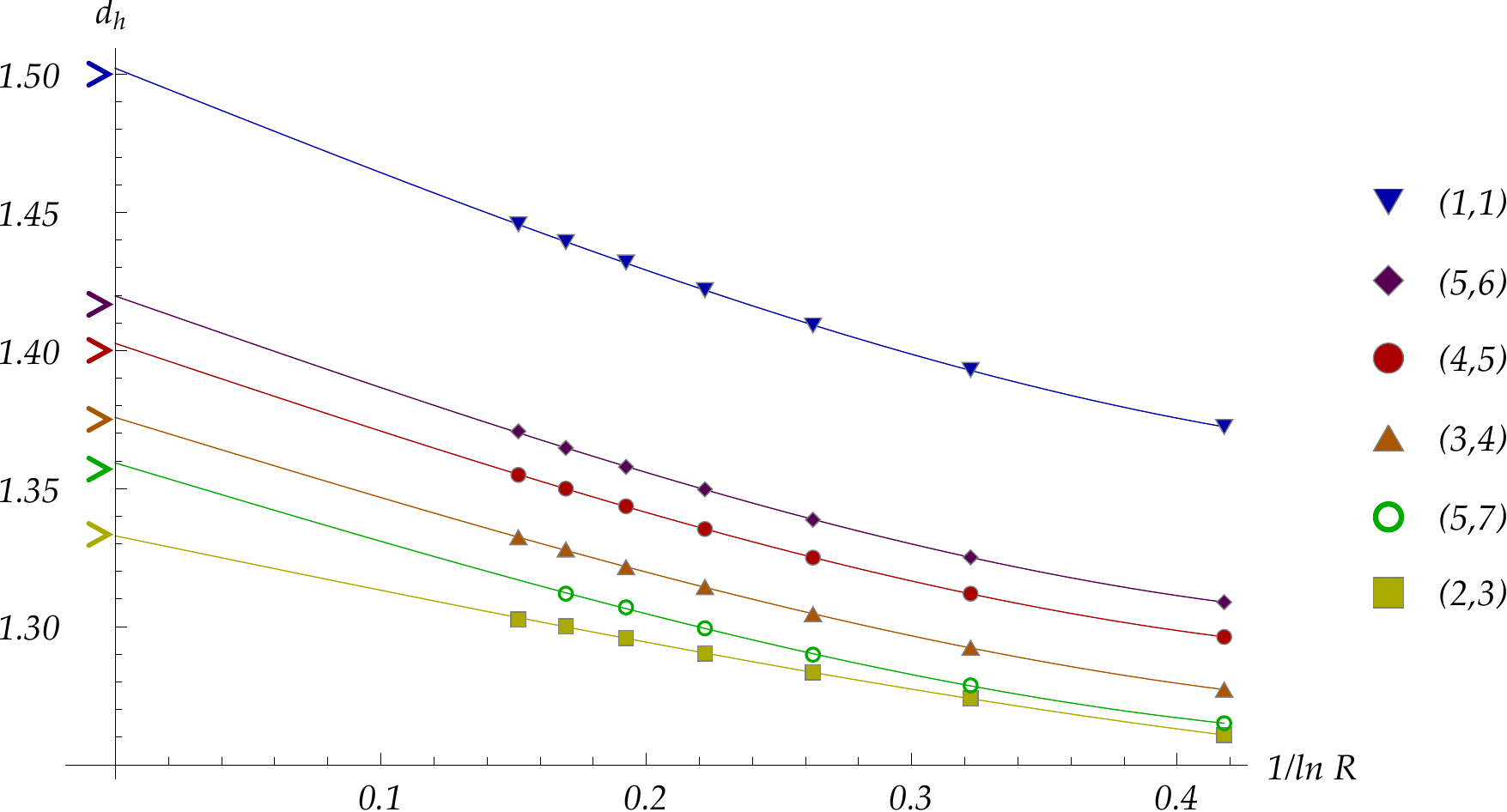}
  \caption{\label{fig:dataH}Measurements of the hull's fractal dimension against $1/\ln R$, and the 
  corresponding fits on the $V/H=2$ cylinder. The theoretical values are designated by wedges at 
  $1/\ln R\to0$. The pair $(p,p')$ refers to $\mathcal{DLM}(p,p')$.}
 \end{center}
\end{figure}

\subsection{The hull}\label{sec:Hull}

The measurements and fits of the hull's fractal dimension are shown in figure~\ref{fig:dataH} for all 
models. The error bars are much smaller than the symbols used to depict 
the data. According to the statistical test used (see appendix~\ref{app:stat}), 
the best fits are obtained for the model $Y(R)=\beta_0+\beta_1/\ln R+\beta_3/\ln^3R$, 
where $\beta_1$ and $\beta_3$ are real constants, $R$ is the linear size of the lattice, 
and $\beta_0$ is a real constant to be 
interpreted as the fractal dimension. Note that the necessity of a quadratic 
term in $1/\ln^2 R$ is always rejected by the statistical analysis. We do not have a straightforward 
explanation for this. The figure underlines the quality of the match between 
theoretical values \eqref{eq:hull} and measured ones and also reveals some of the possible 
difficulties.

First and foremost, the data show that the $\widehat{d_h}$'s, as functions of $1/\ln R$, are not 
linear and have large slopes. The smallest lattice used in the fits has 
$H=8$. On the interval of $H$ covered by our measurements, just a little more than half the 
distance between $\widehat{d_h}^{H=8}$ and the theoretical value is covered. 
More precisely, the ratio of the data span 
$\bigl|\widehat{Y}\bigl(R(512)\bigr)-\widehat{Y}\bigl(R(8)\bigr)\bigr|$ over the distance 
$\bigl|d_h-\widehat{Y}\bigl(R(8)\bigr)\bigr|$, averaged over all fits, is only $0.57$. Even with data 
for larger lattices, say $H=1024, 2048$ and $4096$, this ratio would still be 
under $0.7$. These extrapolations are obviously difficult. Second, the fits are extremely sensitive 
to the non-linear terms of $Y(R)$. According to table \ref{tab:Meas} and 
figure~\ref{fig:dataH}, the fits give an extrapolated fractal dimension that slightly overshoots the 
theoretical one, the only exception being the fit for $\mathcal{DLM}(2,3)$. The fit for 
$\mathcal{DLM}(3,4)$ 
overshoots very slightly, by about $0.0007$, and the others by about $0.002$. The latter are 
precisely those whose graphs are the most curved. The departure from the predicted 
values is even more striking for some values of the external perimeter and we shall propose in the next subsection 
an experiment to better understand the systematic error caused by the extrapolation.

Of course, better estimates of $d_h$ would require data for larger lattices. For our algorithm, the 
major obstacles to reach these lattices are the following. First, the smaller the 
$u$, $v$ and $w$ weights (see \eqref{eq:YBsol}), the slower the simulations. The reason is that, 
when the algorithm steps on an empty $3\times3$ block, the probability that 
the block remains empty ranges from $90$\% for $\mathcal{DLM}(1,1)$ to over $99$\% for 
$\mathcal{DLM}(5,7)$. Because the functions $u$, $v$ and $w$ increase with $\beta$, the 
autocorrelation 
function of the Monte Carlo chain decreases faster for larger $\beta$, as there is less emptiness. 
Second, the dependency of $d_h$ on boundary effects favored simulations on 
the cylinder. But accounting for the cyclical boundary conditions slows the algorithm by about 
$5\%$. Third, the amount of data on the lattices up to $H=256$, for 
$\beta\neq0,1$, are easily stored in the cache of the cpus used but, as $H$ reaches $512$, more 
traffic between the cache and the memory is needed as seen by the sudden 
decrease in the number of Monte Carlo cycles performed per second for these larger $H$
values. Finally, the model $\mathcal{DLM}(5,7)$ is the hardest to measure. As seen in 
figure~\ref{fig:dataH}, $H=256$ is the largest lattice for it. Its fugacity 
$\beta_{(5,7)}\simeq0.6180$ is the lowest among the models measured, excluding 
$\beta_{(2,3)}=0$ that has its own algorithm anyway. The $u$, $v$ and $w$ are small and the 
cost of creating a loop starts to show, leading to rather empty configurations 
with the problem noted above.

\subsection{The external perimeter}\label{sec:EP}

The definition we used for the external perimeter is inspired by the biased walker of 
\citet{Grossman1986} and its variant for loop models introduced in 
\citet{Saint-Aubin2009}. Starting at one end of the cylinder, the walker follows the left side of the 
defect all the way down to the other extremity of the cylinder and, in the 
meanwhile, counts a unit of length each time it encounters a $u$- or $v$-face. When it encounters 
a $w$-face made up of two quarter-circles coming from the defect, it still 
counts a unit of length, but then it does not enter the fjord created at that point, unless the 
two quarter-circles possess azimuthal angles $\varphi_1$ and $\varphi_2$ such that 
$\lvert\varphi_1-\varphi_2\rvert=2\pi$. (See section~\ref{sec:windingNumber} for the definition of 
$\varphi_i$.) That is, if we interpret the defect as a chordal SLE growing from one end of the 
cylinder to the other, then its azimuthal angle 
as it progresses along the trajectory takes the value $\varphi_1$ when it first crosses the $w$-face 
and $\varphi_2$ the second time. The only possible differences of these 
two values are $\lvert\varphi_1-\varphi_2\rvert=0$ and $2\pi$. When the difference is $2\pi$, we 
consider that this $w$-face is not the entry point of a fjord because 
otherwise the walker would return to its starting point without scanning the defect in its entirety. 
The external perimeter for two distinct defects is drawn in figure~\ref{fig:Observables}.

As observed in \citep{Saint-Aubin2009}, there could be more than one definition for the external 
perimeter in loop models. Thus, to confirm the validity of the one chosen, we have
compared the extrapolated value obtained for the dense phase of $\mathcal{DLM}(2,3)$ with that 
of $\mathcal{LM}(2,3)$, previously measured in \citep{Saint-Aubin2009}. 
Since the two models belong to the same universality class, as explained in 
section~\ref{sec:DLM}, the two 
extrapolations should be equal. For $\mathcal{DLM}(2,3)$ with the above 
definition, we measured $1.3368$, while one finds $1.326$ for $\mathcal{LM}(2,3)$ in 
\citep{Saint-Aubin2009}. Both measures are fairly close to each other and to the 
theoretical value of $d_{ep}=4/3$. We thus conclude that there is no reason to doubt our definition.

\begin{figure}[htb!]
 \begin{center}
  \includegraphics[width=0.9\textwidth]{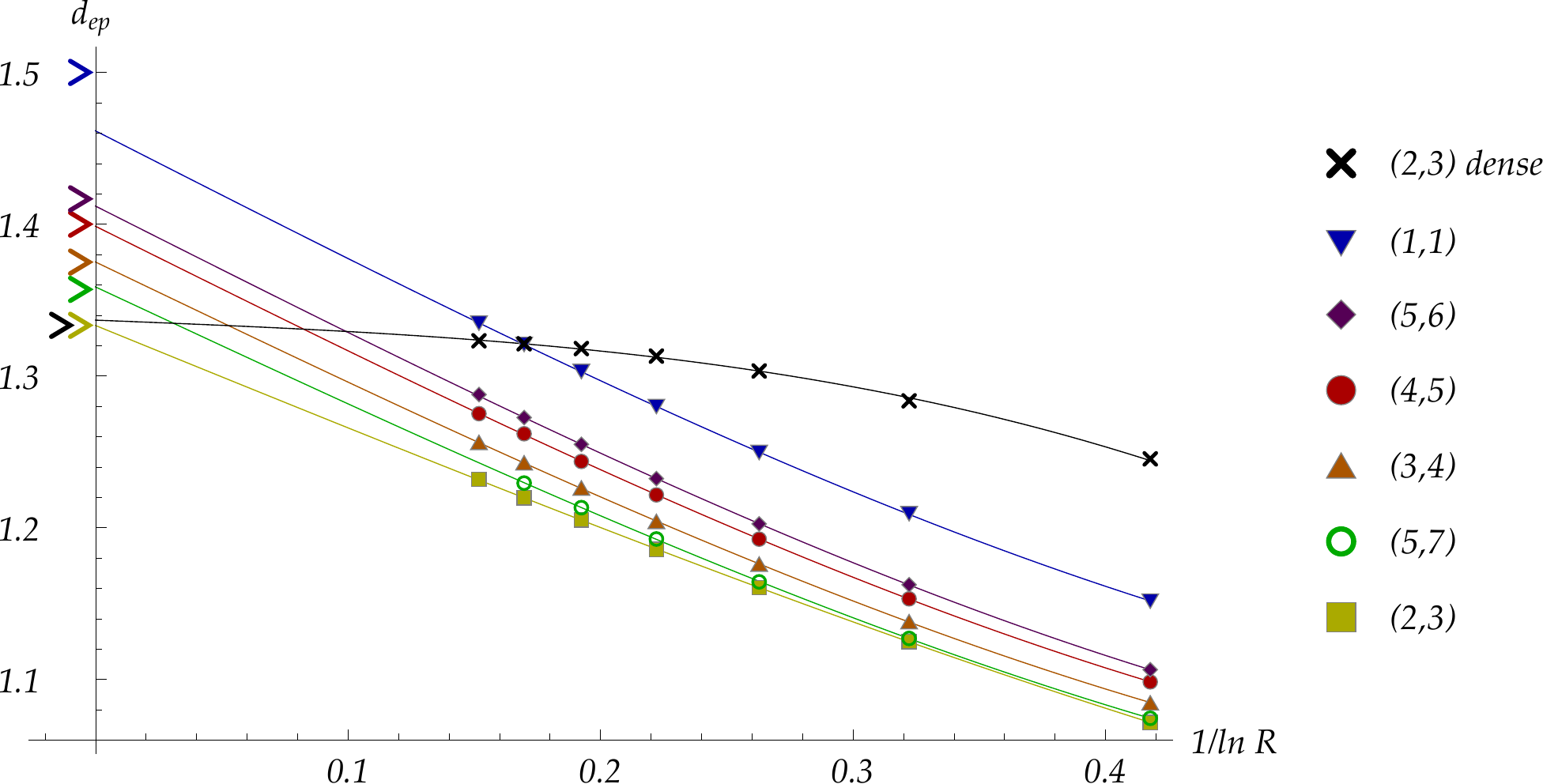}
  \caption{\label{fig:dataEP}Measurements of the external perimeter's fractal dimension, and the 
  corresponding fitted curves, for all the previously cited models on the $V/H=2$ cylinder.
  The pair $(p,p')$ refers to $\mathcal{DLM}(p,p')$.}
 \end{center}
\end{figure}

The measured dimensions $\widehat{d_{ep}}$ for the external perimeter are shown in table 
\ref{tab:Meas}. Again, we see that the agreement with the prediction is generally 
good, although the results for $\mathcal{DLM}(5,6)$, and particularly the one for 
$\mathcal{DLM}(1,1)$, are farther away from 
the theoretical ones than the corresponding results of the hull. Nevertheless, as 
visible in figure~\ref{fig:dataEP}, the fits are good enough to confirm convincingly the theoretical 
predictions of equation \eqref{eq:EP}, except maybe for $\mathcal{DLM}(1,1)$. The latter 
model is in the equivalence class with central charge $c=1$ to which also belongs the $4$-Potts 
model. It is known that, for many models at $c=1$, the scaling of geometric 
objects include logarithmic terms (see for example \citep{Aharony2002}) and thus extrapolation is 
predictably difficult for $\mathcal{DLM}(1,1)$. Still, one would like to better understand the 
systematic undershooting for this geometric exponent and, to a lesser extent, for that of the hull. We 
propose the following experiment with this goal in mind.

\begin{figure}[htb!]
 \begin{center}
  \includegraphics[width=0.85\textwidth]{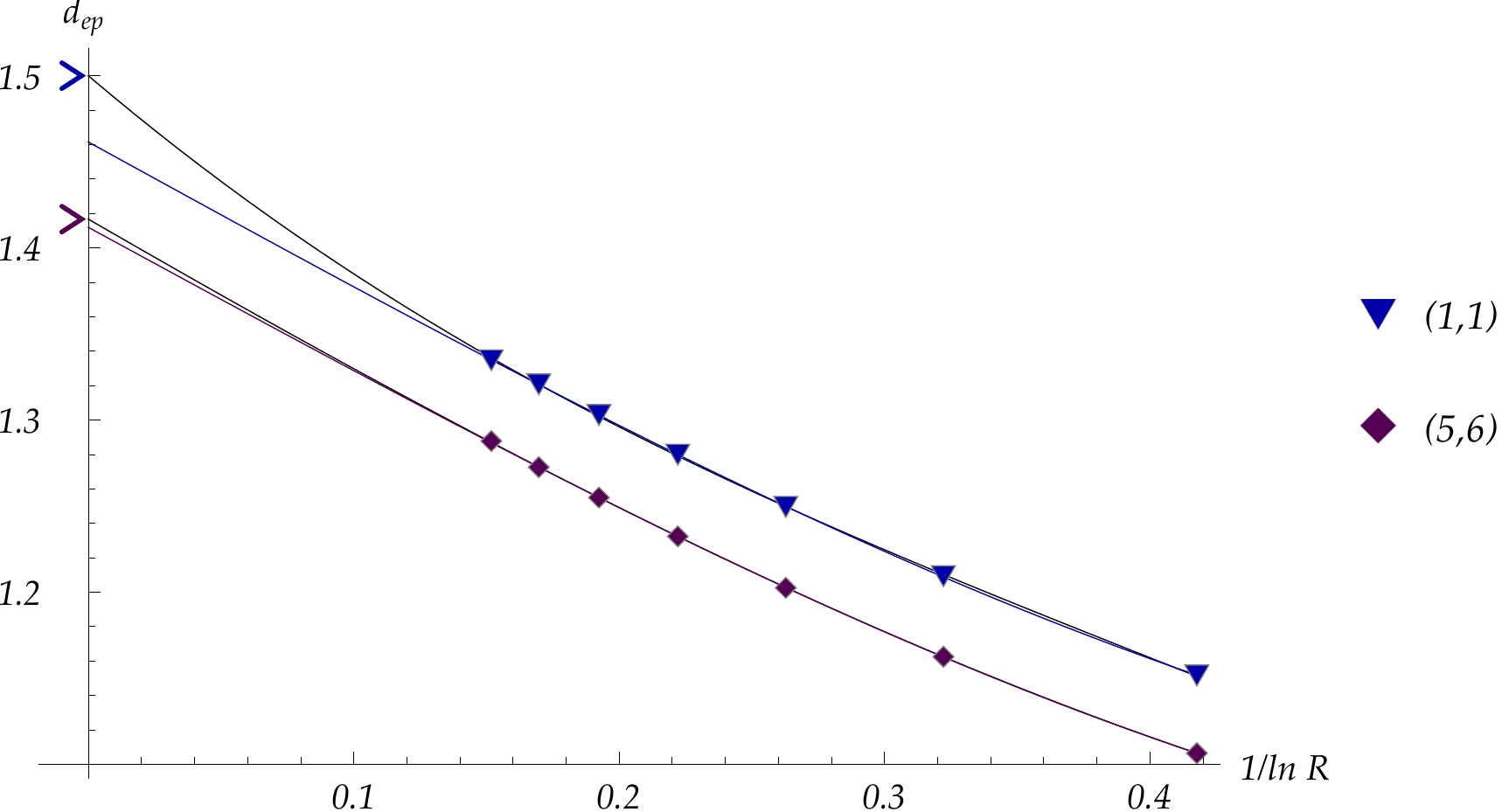}
  \caption{\label{fig:dataEPnew}Fits and measurements for the external perimeter of $\mathcal{DLM}(5,6)$ and 
  $\mathcal{DLM}(1,1)$ as well as the fits with the additional point $d_{ep}^{\text{\rm theo}}$ at $1/\ln R=0$.}
 \end{center}
\end{figure}

To the Monte Carlo data for $\mathcal{DLM}(1,1)$ 
and $\mathcal{DLM}(5,6)$ that we have, we added the point 
$(1/\ln R, d_{ep})=(0,d_{ep}^{\text{\rm theo}})$, i.e.~the theoretical prediction. (This idea is not new. For 
example, \citet{Asikainen2003} also use fits where the theoretical fractal dimensions appear 
explicitly to determine their error bars.) And we fitted the new datasets with this single additional point. According to the $F$-test 
(see appendix~\ref{sec:modelTesting}), it is now appropriate to 
keep both of the terms $1/\ln^2 R$ and $1/\ln^3 R$, and the coefficients of 
$Y(R)=\beta_0+\beta_1/\ln R+\beta_2/\ln^2R+\beta_3/\ln^3R$ are now all significant. Visually the two 
fits, with and without the theoretical value, are barely distinguishable in the 
range of measurements as can be seen in figure~\ref{fig:dataEPnew}, but they split quickly to the 
left of the datum corresponding to the largest lattice. We therefore propose 
the following simple interpretation for the systematic departure from the predicted values: in the 
range $H\le 512$, the cubic term $1/\ln^3 R$ is large enough to conceal the 
quadratic one. Without the additional (predicted) point, the $F$-test says that the hypothesis that 
$\beta_2$ is zero cannot be rejected. (Actually, if one includes $\beta_2$, the 
$\widehat{d_h}$ is similar, but the confidence interval is larger.) The fit with only 
$Y(R)=\beta_0+\beta_1/\ln R+\beta_3/\ln^3R$ is then excellent. But, with the predicted value 
added, the hypothesis that $\beta_2$ is zero must be rejected and the new fit goes through all 
the measured $\widehat{d_h}^{H\times V}$ with equal precision. We made 
one further check by adding, instead of the theoretical value, a set of random points at 
$1/\ln R=0.05$ and $0.10$ whose averages sit on the new fits of 
figure~\ref{fig:dataEPnew} with a variance similar to the spread of the data at $H=512$. 
The fit with these random points had the same properties as the one with the theoretical 
value added. The conclusion seems to be that the sizes of lattice we used do not allow to measure 
properly both $\beta_2$ and $\beta_3$ that describe finite-size corrections 
from $H\simeq 8$ to $H\rightarrow \infty$. The values $1/\ln R=0.05$ and $0.10$ correspond to 
lattices of size $H\simeq 3\times 10^8$ and $2\times 10^4$. They are out of 
reach with our algorithms!

\subsection{The red bonds}\label{sec:redBonds}

\begin{table}[htb!]
 \begin{center}
  \begin{tabular}{|c||c|c|c|c|c|c|}
   \hline 					&					&					&					&					&					&					\\[-2mm]
   $H$						&	$8$				&	$16$				&	$32$				&	$64$				&	$128$				&	$256$				\\[2mm]
   \hline 					&					&					&					&					&					&					\\[-2mm]
   $\widehat{d_{rb}}$			&	$0.1315\bigl|\bigr.5$	&	$-0.0243\bigl|\bigr.4$	&	$-0.1262\bigl|\bigr.7$	&	$-0.1991\bigl|\bigr.9$	&	$-0.2510\bigl|\bigr.11$	&	$-0.2941\bigl|\bigr.21$	\\[2mm]\hline
  \end{tabular}
 \end{center}
 \caption{\label{tab:RB}The measured fractal dimension of red bonds for the Ising model 
 $\mathcal{DLM}(3,4)$ on cylinders \mbox{$H={8,16,32,64,128,256}$} with $V/H=2$.}
\end{table}
To measure the fractal dimension of red bonds, we use two defects, instead of just one, so that the 
domain between them is interpreted as a proper ``cluster''. The fractal 
dimension is then computed by finding the number of $w$-faces visited by both 
defects and multiplying this number by two (see section~\ref{sec:geoFractalDim}). The 
measurement of this 
observable is difficult since, for models in the dilute phase, not even all configurations have at 
least a single red bond. Moreover, the average number of red bonds is smaller 
than $1$. Because of this, the fractal dimension cannot be estimated using equation 
\eqref{eq:avValue} and we used an alternative definition that was also used in 
\citet{Saint-Aubin2009}:
\begin{equation}
 \label{eq:avValueRB}
 \widehat{d_{rb}}=\frac{1}{n}\sum_{j=1}^n\frac{\ln\overline{N_j}}{\ln R(H)},
\end{equation}
where $\overline{N_j}=\frac{1}{m}\sum_{i=1}^mN_{i,j}$ and $N_{i,j}$ is the number of red bonds of 
the $i$-th datum in the $j$-th Markov chain. The numbers $m$, $n$ and $R(H)$ have the 
same meaning as in section~\ref{sec:Hull}.

As the data of table \ref{tab:RB} attest, there are less and less red bonds in average for each 
lattice configuration as $H,V\to\infty$. Because of this, we restricted our 
measurements to $\mathcal{DLM}(3,4)$, as it has the most efficient algorithm, and 
confined the lattices to $H\in[8,256]$. The difficulty is well illustrated by the 
following numbers:  the average value $-0.2941\pm0.0021$ at $H=256$ in table \ref{tab:RB} 
required a total of about $10^{15}$ Monte Carlo iterations, excluding the 
number of iterations needed for an adequate thermalization of the startup lattices. 
By comparison, the result 
$1.3323\pm0.0004$ at $H=512$ for the hull required about the same 
number. In other words, obtaining a measurement for the hull at $H=256$ with a similar 
confidence interval of $\pm0.002$ would require only about $5.5\times10^{12}$ 
iterations, i.e.~approximately $1/180$ times the number required in the red bonds experiment.

\begin{figure}[htb!]
 \begin{center}
  \includegraphics[width=0.8\textwidth]{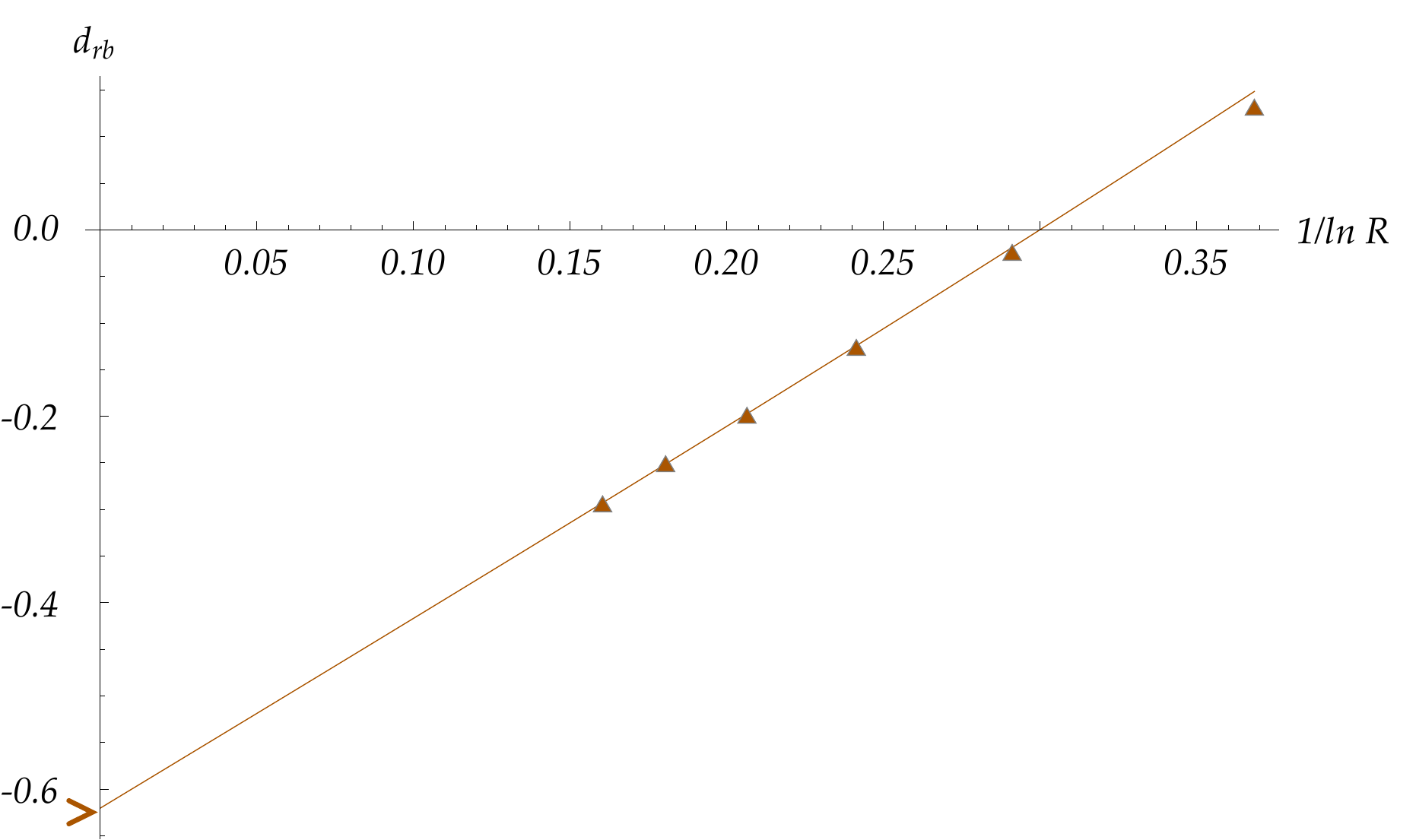}
  \caption{\label{fig:dataRB}Measurements of the red bonds' fractal dimension and its 
  corresponding fit, for the dilute Ising model $\mathcal{DLM}(3,4)$ on the $V/H=2$ 
  cylinder.}
 \end{center}
\end{figure}

Although some of the measurements are less precise than for the other observables, we obtained 
the extrapolated $\widehat{d_{rb}}=-0.621\pm0.004$, in good agreement 
with the predicted value $d_{rb}=-0.625$ (eq.~\eqref{eq:redBonds}). This result is surprisingly 
close to the theoretical value, considering the extreme extrapolation that 
had to be done: the value at $H=256$ remains at about $0.33$ from the theoretical prediction, that 
is $10$ times farther than the corresponding value was for $d_{ep}$ in 
$\mathcal{DLM}(1,1)$. However, the fit gives a good result because the curvature is 
gentler, and the discrepancy from linearity is rapidly wearing off as $1/\ln R$ 
decreases. The average values as well as the fit for this observable have been plotted in 
figure~\ref{fig:dataRB}.

The {\em negative} fractal dimension $d_{rb}$ is somewhat of a novelty here, but can be easily 
interpreted. As the lattice size grows, the average number of times the two 
defects touch one another decreases toward zero, and the rate of this decrease is measured by 
the negative fractal dimension $d_{rb}$.


\section{Concluding remarks}

The agreement between the theoretical values and the measured ones (table \ref{tab:Meas}) is 
very good. It is similar in precision to that obtained in \citep{Saint-Aubin2009} for the dense 
models, or even slightly better. Note that the difficulties that had to be overcome for the dilute 
models were quite different from those encountered for the dense ones. The improved
quality of the geometric exponents obtained here might stem partially from a more systematic and 
finer statistical analysis (see appendix~\ref{sec:Stats}). 

The proposed upgrade algorithm for the dilute models, and its variants for the models with 
$\beta=0$ and $1$, was quick enough to provide very precise measurements for 
lattice sizes up to $H\times V=512\times 1024$. Still, one feels that, unless willing to wait for the 
next or the second next generation of cpus, the algorithm proposed here has 
reached its practical limits. New methods, in the direction of those proposed by \citet{Deng2007}, 
will be necessary to probe further these models.


\section*{Acknowledgments}
This work is supported by the Canadian Natural Sciences and Engineering Research Council 
(Y.S.-A.) and the Australian Research Council (P.A.P.~and J.R.).
J.R. is supported by the Australian Research Council under the Future Fellowship
scheme, project number FT100100774.
We acknowledge Amelia Brennan who carried out some preliminary numerical calculations
for small lattices as part of her Summer Vacation Scholarship at Melbourne University.


\appendix


\section{Upgrade algorithms}

In this appendix, we first present the upgrade algorithm common to all the models. The 
subsequent subsections sketch the ideas relevant for particular models.

\subsection{The basic algorithm}\label{sec:Algorithm}

For the dense loop models $\mathcal{LM}(p,p')$, reviewed in appendix~\ref{sec:LM}, 
each box is in either one of two faces, corresponding to the two $w$-faces of 
$\mathcal{DLM}(p,p')$. A straightforward Metropolis-Hastings upgrade can be chosen as simply 
changing one box at a time and checking whether the usual Monte-Carlo 
condition is verified. (See \citep{Saint-Aubin2009}.) For the dilute loop models 
$\mathcal{DLM}(p,p')$, the edges of the nine possible states are not necessarily crossed by loop
segments, as is shown in figure~\ref{fig:Weights}. Indeed, the empty state contains no loop
segments, two edges of the 
$u$- and $v$-faces are crossed by a loop segment and all four edges of the $w$-faces are 
crossed. As the loop segment must remain continuous at box interfaces, 
we could refine the previous algorithm by 
requiring that the edges that are crossed before the change, and only those, 
remain crossed after the change. But then the only effective change would be the flipping of a 
$w$-face to its mirror face, thus preventing the algorithm from sampling over all 
configuration space. Because of this, we are forced to consider changing many contiguous boxes 
in a single upgrade step.

\begin{figure}[h!]
 \begin{center}
  \subfigure[The chosen block.]{\label{fig:3x3Chosen}\includegraphics[width=0.2\textwidth]{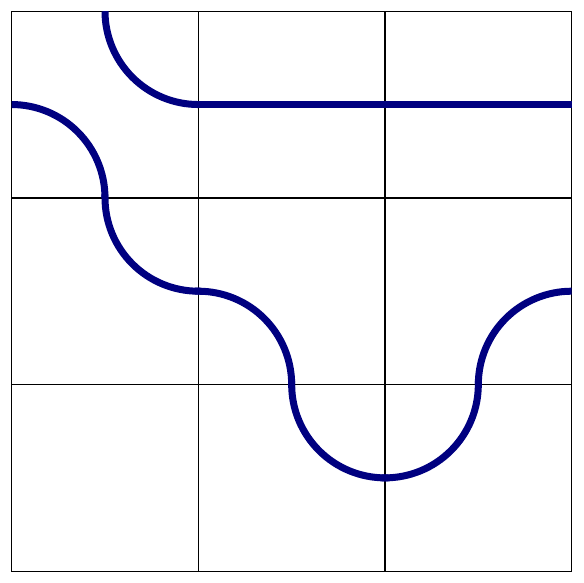}}\hfill
  \subfigure[An admissible replacement block.]{\label{fig:3x3Works}\includegraphics[width=0.2\textwidth]{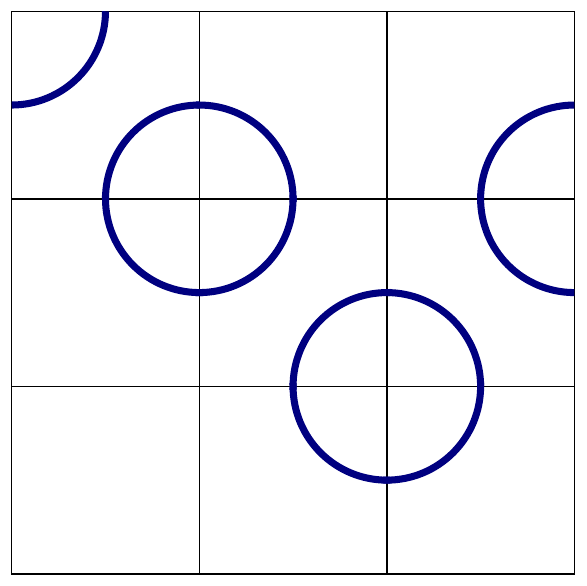}}\hfill
  \subfigure[A forbidden replacement block.]{\label{fig:3x3DoesntWork}\includegraphics[width=0.2\textwidth]{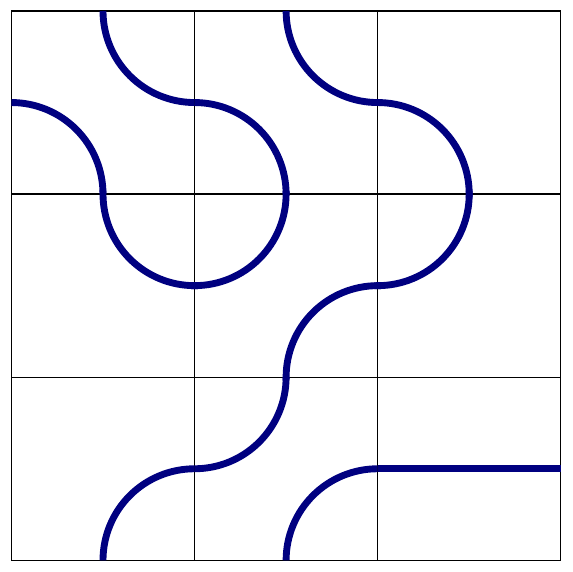}}
  \caption{\label{fig:Change3x3}The original $3\times 3$ block in (a) with an admissible replacement in (b) and a forbidden one in (c).}
 \end{center}
\end{figure}

The upgrade step must therefore change an $m\times n$ block of boxes, with $m,n\ge 2$. First an 
$m\times n$ block is chosen at random in the $H\times V$ lattice on the 
cylinder. Because this block has to fit entirely in the lattice, this amounts to placing randomly 
the upper left box of the block in the first $V-n+1$ rows, all possibilities 
being weighted uniformly. Second, the content of the 
$m\times n$ block is changed for another \emph{admissible} block. To be 
admissible, the $2(m+n)$ edges of the block must be crossed by a loop if they were in the original 
block, and be free of crossing if the original edge had none. For instance, if 
the chosen $3\times 3$ block corresponds to figure~\ref{fig:3x3Chosen}, then an admissible 
replacement is shown in figure~\ref{fig:3x3Works}, while 
figure~\ref{fig:3x3DoesntWork} shows a forbidden replacement block. 
The new block must be chosen uniformly among admissible ones.
Finally, the Boltzmann weight $u^{n_u}v^{n_v}w^{n_w}\beta^N$ of 
the configuration where the $m\times n$ block is replaced by the new choice is computed and 
compared with the original one (see \eqref{eq:OnSquareLoop}): the Metropolis-Hastings ratio decides whether 
the replacement is to be accepted or rejected. Note that, even though the change is limited to the 
block, the computation of the weight involves counting the number of closed 
loops (except when $\beta=1$) and might therefore require exploring the configuration at a large 
distance of the $m\times n$ block under consideration.

The requirement of uniformity of the block state among admissible ones raises a subtle problem. 
One might think that it is achieved by simply choosing the face of each of the 
$m\times n$ boxes one after the other, respecting at each step the conditions on the perimeter. 
This is not the case as the next example shows. Suppose that the $m=n=3$ block 
to be changed is that of figure~\ref{fig:3x3Chosen} and that the new box states will be chosen from 
left to right, top to bottom. If the first box to be chosen is the upper left one, 
three choices are admissible: the $u$-face $u_2$ that has a single west-north crossing, and both 
$w$-faces $w_1$ and $w_2$. (As in section \ref{sec:squareLattice}, the index of the letters $u$, $v$ and $w$ 
refers to the order of figure~\ref{fig:Weights}, the bottom box being labeled by $1$.) Each of these 
three faces will be given probability $\frac{1}{3}$. It is easy to check 
that there are two possible replacements for the next box, and only one for the last box of the top 
row. Similarly, two faces are admissible for the leftmost box 
of the second row. So far, twelve different fillings are possible, each occurring with probability 
$\frac{1}{12}$.
\begin{figure}[t]
 \begin{center}
  \subfigure{\label{fig:moreProbable}\includegraphics[width=0.2\textwidth]{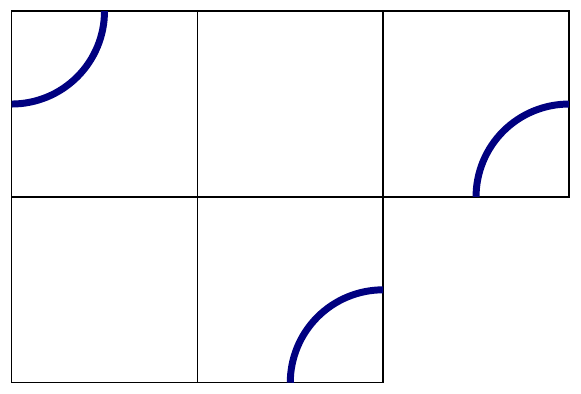}}\hspace{3cm}
  \subfigure{\label{fig:lessProbable}\includegraphics[width=0.2\textwidth]{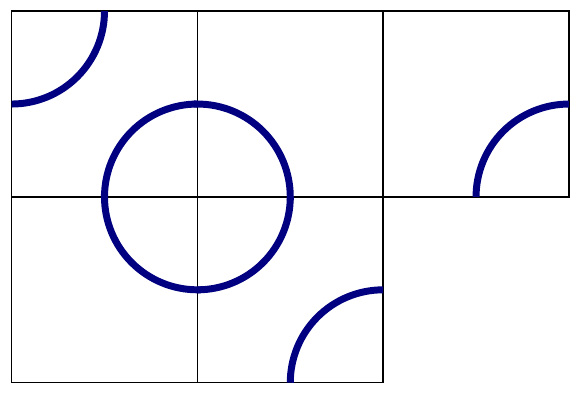}}
  \caption{\label{fig:partialFillings}If we chose the face of each box sequentially, then the left block would be more probable than the right one.}
 \end{center}
\end{figure}
The difficulty occurs for the box at the center. Two of the twenty-seven possible fillings are shown 
in figure~\ref{fig:partialFillings}. For the filling on the left, two choices are 
allowed: the $u_4$ or the empty face. But for the filling on the right, three are possible, namely the 
$u_2$, $w_1$ and $w_2$ faces. This means that, starting from the center box, the 
blocks obtained from the filling on the left will occur with probability $\frac{1}{24}$, and the ones 
obtained from the right filling will get probability $\frac{1}{36}$: this violates 
the requirement of uniformity. To assure uniformity, one has to determine first, for a given block, 
the number of allowed replacements. We found it more efficient to count 
beforehand these for all the possible edge configurations of the perimeter, or boundary state, for 
the $m\times n$ block, and actually construct a list of the admissible 
replacements. For $m=n=3$, there is a total of $113\,361$ possible replacements for the $2048$ 
boundary states. Depending on the boundary state, there can be as little as 
$18$ admissible replacements or as many as $690$.

We decided to work with $3\times 3$ blocks. The list of possible replacements for larger blocks 
would take up much more memory. Moreover, given some boundary state on 
the block being changed, the variations in probability brought by tentative replacements would 
likely be in a wider range and lead to a higher rejection rate. Such an 
argument does not hold for $2\times 2$ blocks, the smallest that allow exploration of the whole 
configuration space, for which the list of replacements is short and the 
acceptance rate is high. However, some experimentation with this shorter list shows that the 
autocorrelation between upgrades, and therefore the number of upgrades between 
statistically independent measurements, is very high. We found the $3\times 3$ blocks to be a 
good compromise.

\subsection{Improvements}\label{sec:Improvements}

The Ising model is described by the dilute phase of $\mathcal{DLM}(3,4)$. Since the loop fugacity 
of this model is $\beta=1$, there is no need to know the number of loops 
crossing the $m\times n$ block when computing the Boltzmann weight. This simplification greatly 
speeds up the algorithm because loop counting is by far its most 
time-consuming component.

It is more difficult to improve the algorithm for models with $\beta\in(0,2]\setminus\{1\}$. The rest of 
this paragraph is devoted to this problem, while the case $\beta=0$ will 
be discussed in subsection~\ref{sec:beta0}. One can distinguish between models according to 
whether $0<\beta<1$ or $1<\beta\leq2$, because models of the first type will 
tend to be filled with less loops than those of the second type. Most configurations of both types of models have large loops, some being of length of the same order as 
that of the defect.

Our first attempt to compute Boltzmann weights was rather naive. We simply followed every 
connected path that crosses the $m\times n$ block until it returns to its starting 
point or, if the path is part of the defect, reaches the boundary of the lattice; that works, but it is 
slow, especially for large lattices. It is slow due to the presence of the defect and,
potentially, of long loops. The problem is particularly acute for those models with large $\beta$, 
e.g.~the dilute phase of $\mathcal{DLM}(1,1)$, because the predicted hull 
fractal dimension of the defect is large and lots of loops are 
present.

A way of reducing the impact of the defect crossing the $m\times n$ block is to assign a time order 
to each edge it crosses, starting from its entry point down to its exiting 
point. Because every edge crossed by the defect is then identified in some way as belonging to it, 
it is possible to distinguish the defect from a loop during loop counting. 
Moreover, the time ordering allows to identify when the defect enters first the block and when it 
leaves it for good. Since these two edges cannot change when going from the 
original to the replacement block, it allows for quick identification of those situations where the 
defect generates a loop or absorbs one intersecting the block.

Large loops are common, some with size commensurate to that of the defect. When the 
$m\times n$ block selected for the upgrade is crossed by one of these, a slowing 
down of the algorithm similar to that encountered for the defect is observed. To get over this 
problem, we time-ordered the edges of each loop, starting from an arbitrary 
edge on its path, and also assigned a unique number to each loop. During the loop counting 
phase of the former block, the number of different loops crossing the block is 
now obtained very quickly: it is simply the total of different loop numbers. The time order is useful 
because it allows the algorithm to know how the different edges of the same 
loop or defect are connected to each other \emph{outside} the $m\times n$ block. Like for the entry 
and exit points of the defect, those \emph{external connections} will not 
change after replacement of the $m\times n$ block. So, with this information, during the loop 
counting step for the replacement block, the algorithm does not have to follow the 
path of the loops (or defect) which lies outside the block since the reentry point is now known.

\subsection{The case $\beta=0$}\label{sec:beta0}

\begin{figure}[htb!]
 \begin{center}
  \subfigure[The block to replace.]{\label{fig:beta0a}\includegraphics[width=0.3\textwidth]{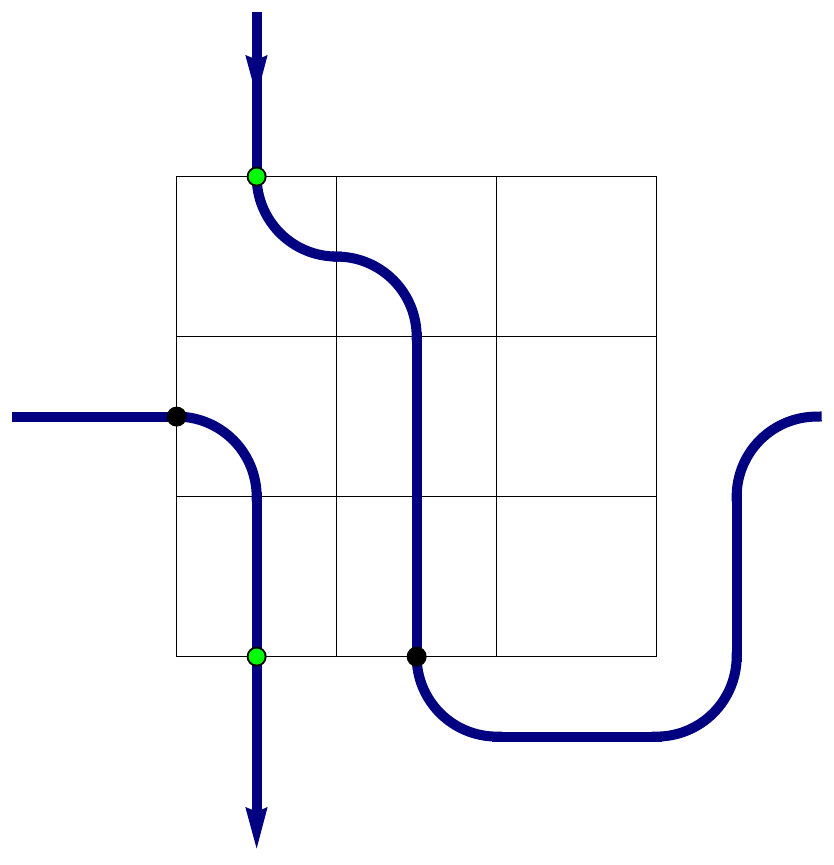}}\hspace{3cm}
  \subfigure[An admissible replacement.]{\label{fig:beta0b}\includegraphics[width=0.3\textwidth]{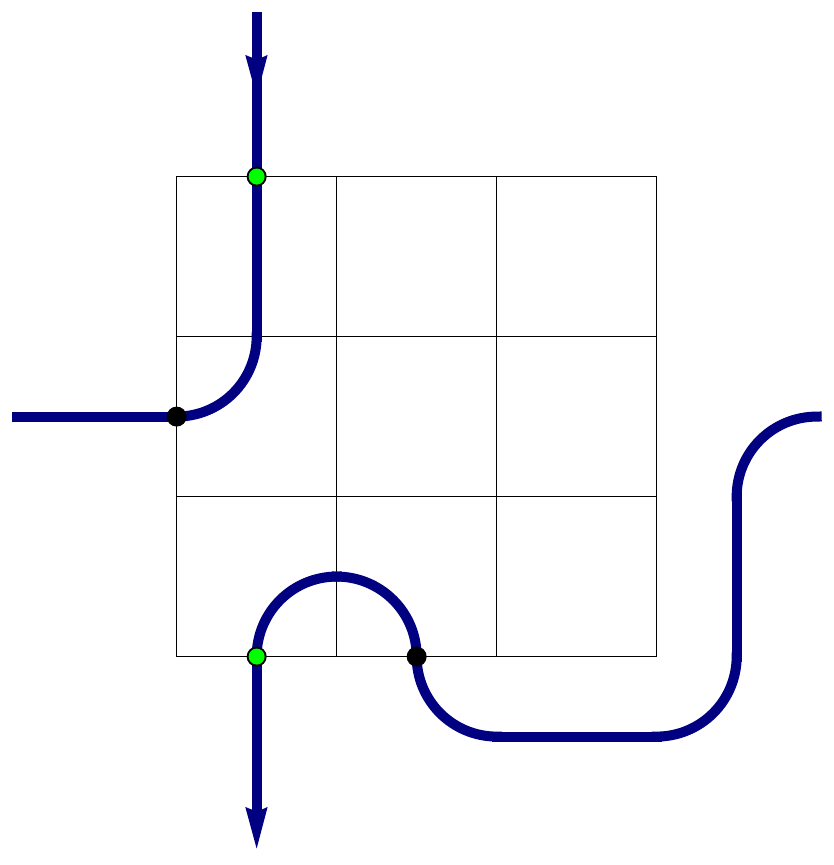}}
  \caption{\label{fig:beta0}A possible transition for a $3\times3$ block on the cylinder yielding no 
  loop. The first and last boundary edges crossed by the defect are marked by pale dots, while the 
  remaining connected edges are marked by black ones.}
 \end{center}
\end{figure}

The percolation model, as described by the dilute phase of $\mathcal{DLM}(2,3)$, needs a special 
algorithm to be simulated efficiently since its $\beta=0$ loop fugacity forbids 
the presence of loops. It is still possible to choose an $m\times n$ replacement block on the sole 
basis that it suits the boundary conditions of the block, but then if the chosen 
configuration creates a loop, this choice will have to be rejected. To get rid of this problem, we 
need to know what the external connections of the block are, as described in
section~\ref{sec:Improvements}; this information is easily retrieved if the defect is 
time-ordered. All that is left to do is to choose, with uniform probability, a 
replacement amongst the blocks respecting both the boundary conditions \emph{and} external 
connections. In figure~\ref{fig:beta0}, an example of a possible replacement 
admitting no loop is shown for a $3\times3$ block on a lattice with cylindrical geometry.

\begin{figure}[thb!]
 \begin{center}
  \includegraphics[width=0.25\textwidth]{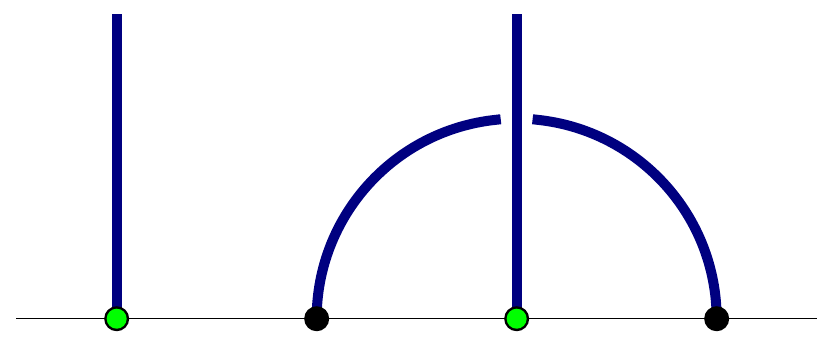}
  \caption{\label{fig:beta0diag}Diagrammatic representation of the external connections shown in figure \ref{fig:beta0}. The points are aligned by starting from the top left edge and 
  following the boundary clockwise.}
 \end{center}
\end{figure}

To obtain a new list of replacement blocks appropriate for this model, the external connections of 
the $m\times n$ block have to be taken into account, so it is 
necessary to be able to enumerate them all. To do so, one may use a diagrammatic 
representation, an example of which can be seen in figure~\ref{fig:beta0diag}. This 
example of a diagram corresponds to the external connections of figures~\ref{fig:beta0a} and 
\ref{fig:beta0b}. Under this representation, the connected edges are linked 
pairwise by arches, and the entry and exit edges of the defect in the block, depicted by pale dots in 
the figures, are linked at infinity by vertical lines. The linking 
arches may not cross each other, but they may go ``under'' the vertical lines, that is without 
intersecting them, as the defect may wind around the cylinder without 
self-intersecting. For given block boundary conditions, consisting of $p$ connected and 
$2m+2n-p$ unconnected edges, there are 
\begin{equation}
 \label{eq:nbEC}
 E_p=\frac12p(p-1)C_{p-2},
\end{equation}
different external connections, where $C_i=\binom{i}{i/2}-\binom{i}{i/2-1}$. To obtain the new list of 
replacements, one must first fix the boundary conditions and external 
connections of the $m\times n$ block and then, using the list of section~\ref{sec:Algorithm}, check 
whether any of the suggested replacements yields loops; those that do not generate 
loops belong to the new list. This testing has to be repeated for all possible external connections 
matching these fixed boundary conditions. Finally, repeating this process for all 
$2048$ boundary conditions completes the list of replacements. For a $3\times3$ block, we find, 
using equation \eqref{eq:nbEC}, a total of
$$
 N=\sum_{\substack{i=2 \\ i\;\text{even}}}^{12}\binom{12}{i}E_i=144\,408
$$
different external connections, leading to a list of over $3.8\times10^6$ replacements. This is to be 
compared with the $113\,361$ possible replacements of the standard 
$\beta>0$ list.

When the $\beta=0$ algorithm steps on an empty $m\times n$ block, nothing can be done since 
loops are not allowed, and so the algorithm skips this block and chooses a new 
one. According to equation \eqref{eq:YBsol}, the $\beta=0$, or $\lambda=\pi/8$, model is the most 
diluted of all the $\beta\geq0$ models. That is, the ratios of the $u,v$ 
and $w$ weights with that of the empty face are at their smallest. Empty blocks therefore occur 
frequently for $\beta=0$ and the present algorithm takes advantage of this. 
Even though the $\beta=0$ algorithm is involved and its list of replacements is heavy, it is the 
second fastest algorithm, second only to the one for $\beta=1$. By second 
fastest we mean that the algorithm for $\beta=0$ computes in average the second most 
Metropolis-Hastings iterations per second.


\section{Statistics}\label{sec:Stats}

In this section, we explain the procedures we used to obtain warm-up intervals, the statistical 
formulas we used to obtain the confidence intervals on measurements, and the 
extrapolation procedure.

\subsection{Warm-up interval}

The \emph{warm-up interval}, or \emph{burn-in period}, is the average number of Monte Carlo 
cycles needed to attain thermalization. To find a reliable warm-up interval for 
each of the models $\mathcal{DLM}(p,p')$, one may use the standard procedure 
(see \citet{Fishman2006} for instance). From any $H\times V$ lattice configuration, start the 
Monte Carlo algorithm for $n$ independent Markov chains and take $m$ measures of the
length $L$ of the defect in the bulk. Each measure is to 
be separated from the next one by $\Delta$ Metropolis-Hastings (MH) cycles. Finally, plot the 
measurements, averaged on these $n$ chains, against $m$: the abscissa where the average 
values stabilize provides a warm-up interval. 

\begin{figure}[htb!]
 \begin{center}
  \includegraphics[width=0.8\textwidth]{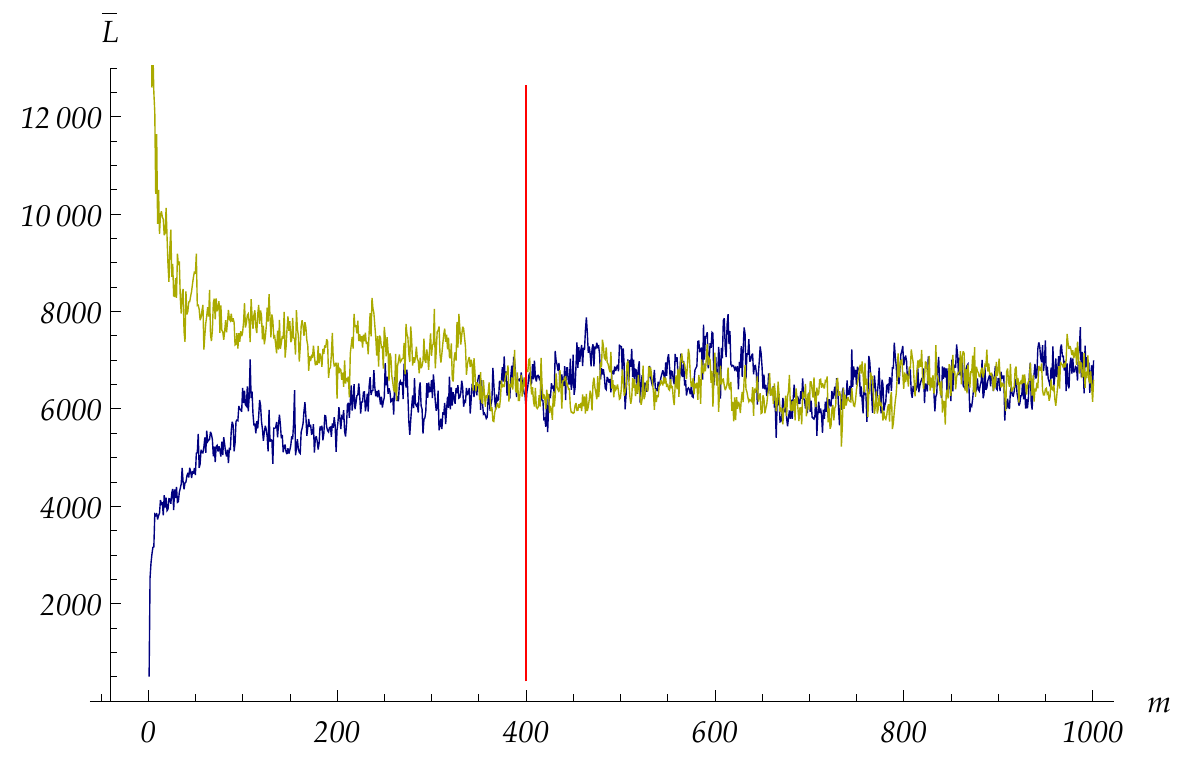}
  \caption[Warm-up interval for $\mathcal{DLM}(3,4)$ $512\times1024$.]{\label{fig:therm34}
  Average defect length $\overline{L}$ in the bulk as a function of the number of measures for a 
  $\mathcal{DLM}(3,4)$ $512\times1024$ lattice on the $V/H=2$ cylinder. This plot was realized 
  for $n=20$, $m=1000$ and $\Delta=200\times10^6$.}
 \end{center}
\end{figure}
The approximate point of thermalization for $\mathcal{DLM}(3,4)$ on the cylinder of size $H=512$ 
and $ V=1024$ is represented in figure~\ref{fig:therm34} by a vertical 
line at $m=400$. After that point, the average value fluctuates around $\overline{L}=6500$ which, 
according to equation \eqref{eq:Dfasym} with $R=\sqrt{523\,776}\simeq723.7$, 
corresponds to $d_h^{H\times V}=1.33$. This is to be compared with the final measurement of 
table \ref{tab:Meas34}, $\widehat{d_h}^{H\times V}=1.3323\pm0.0004$. To 
illustrate the reliability of this thermalization point, we plotted two datasets in 
figure~\ref{fig:therm34}: the dark one corresponds to a set of measures started from a 
configuration having the smallest possible defect length in the bulk, that is 
$L=L_{\text{min}}=512$, while the light one corresponds to a set started from the maximal defect 
length $L=L_{\text{max}}=R^2=523\,776$.

\subsection{Statistical analysis}\label{app:stat}

We give the details here of the methods used for obtaining confidence intervals on the different 
measures, on how the linear regressions, or fits, of the data were done, and 
also what processes were used to discriminate between a good and a bad linear regression. For 
more details on these subjects, see \citet{DS1998}.

\subsubsection{Confidence interval}\label{sec:confInterval}

For an experiment consisting of $n$ independent Markov chains, e.g.~$n$ computers or 
processes, and a total of $m$ measurements $Q_{i,j}$ per chain, where $i$ 
labels the chain and $j$ the datum, the unbiased estimator of the expected value 
$\mathrm{\mathbf{E}}[Q]=\overline{Q}$ of an observable $Q$ is given by the grand sample 
average
$$
 \widehat{Q}=\frac{1}{n}\sum_{i=1}^{n}\overline{Q}_i,
$$
where $\overline{Q}_i=\frac{1}{m}\sum_{j=1}^{m}Q_{i,j}$ is the average value of the $i$-th Markov 
chain. To obtain a confidence interval on $\widehat{Q}$, we first need 
to compute the unbiased sample variance
$$
 \widehat{\sigma}^2=\frac{1}{n-1}\sum_{i=1}^{n}\left(\overline{Q}_i-\widehat{Q}\right)^2.
$$
In this work, some of the observables $d_S^{H\times V}$ were measured with a small number $n$ 
of Markov chains ($n\sim20$). In these cases, it is better to replace the 
approximate $95\%$ confidence interval $2\widehat{\sigma}/\sqrt{n}$ by the correct
\begin{equation}
 \label{eq:CI}
 \left[\widehat{Q}-\tau_{n-1}(0.95)\frac{\widehat{\sigma}}{\sqrt{n}},\;\widehat{Q}+\tau_{n-1}(0.95)\frac{\widehat{\sigma}}{\sqrt{n}}\right],
\end{equation}
where $\tau_{n-1}$ is the inverse Student t-distribution with $n-1$ degrees of freedom. For 
$n=20$ observations, the factor $\tau_{19}$ of equation \eqref{eq:CI} equals 
approximately $2.093$.

\subsubsection{Linear regression}\label{sec:linReg}

The estimated fractal dimension of the different observables in the continuum scaling limit, as
$R\rightarrow\infty$ in \eqref{eq:Dfasym}, is obtained by extrapolating the 
linear regression of the measurement data acquired at finite $R$'s. For an experiment of 
$n_1+n_2+\dotsb+n_\ell\equiv n$ total measures (sample points) of the observable, 
with $n_i$ the number of measures for the $i$-th lattice size (e.g. $H_i=32,H_{i+1}=64$, etc.), the 
most general linear model with $p$ parameters is:
\begin{equation}
 \label{eq:statModeli}
 Y_i=\beta_0+\beta_1X_{i1}+\beta_2X_{i2}+\dotsb+\beta_{(p-1)}X_{i(p-1)}+\epsilon_i,
\end{equation}
where $Y_i$, $i=1,2,\dotsc,n$, are the sample points, $\beta_k$, $k=0,1,\dotsc,p-1$, are 
parameters associated with the $p$ independent variables $X_{ik}$, and 
$\epsilon_i\sim N\left(0,\sigma_i^2\right)$ is the $i$-th random error associated with $Y_i$. We 
shall be interested in polynomial fits, for which $X_{ij}=X_i^j$, with 
$X_i=1/\ln R(H_i,V_i)$. Moreover, $Y_i$ corresponds to the measured fractal dimension 
$d_S^ {H_i\times V_i}$ at $R(H_i,V_i)$. The equation \eqref{eq:statModeli} can be 
rewritten in matrix notation as
\begin{equation}
 \label{eq:statModel}
 \mathbf{Y}=X\boldsymbol{\beta}+\boldsymbol{\epsilon},
\end{equation}
where $\mathbf{Y}$ and $\boldsymbol{\epsilon}$ are $n\times1$ vectors, $\boldsymbol{\beta}$ is 
a $p\times1$ vector, and $X$ is a $n\times p$ matrix. Note that the first 
column of $X$, related to $\beta_0$, is filled with $1$'s. The fitted equation,
\begin{equation}
 \label{eq:linReg}
 \widehat{\mathbf{Y}}=X\boldsymbol{\widehat{\beta}},
\end{equation}
is such that $\widehat{\mathbf{Y}}=\mathrm{\mathbf{E}}[\mathbf{Y}]$ and 
$\boldsymbol{\widehat{\beta}}=\mathrm{\mathbf{E}}[\boldsymbol{\beta}]$ are unbiased 
estimators of $\mathbf{Y}$ and  $\boldsymbol{\beta}$, respectively; these are the quantities we 
need to evaluate.

The difficulty with the present datasets is that the observables $d_S^{H_i\times V_i}$ have not 
been measured with the same precision, that is, their variances depend on the 
size of the lattice. And when 
$\mathrm{\mathbf{Var}}[\boldsymbol{\epsilon}]\equiv\sigma^2V\neq\sigma^2\mathds{1}_n$, with 
$\sigma^2V_{ij}=\mathrm{\mathbf{Cov}}[\epsilon_i,\epsilon_j]$ and $\sigma$ a positive constant, 
one cannot rely on \emph{ordinary least squares} (OLS) for obtaining an 
unbiased linear regression, but must count instead on \emph{weighted least squares} (WLS). In 
other words, if the sample variances 
$\sigma^2V_{ii}=\mathrm{\mathbf{Var}}[\epsilon_i]=\mathrm{\mathbf{Var}}[Y_i]$ in 
\eqref{eq:statModeli} cannot be considered constant throughout the dataset, one 
cannot use OLS. Moreover, the WLS method may be used only if $V$ is diagonal, i.e. if the 
different $\epsilon_i$ are uncorrelated. Fortunately, this is the case here as our 
Markov chains are independent.

The idea behind WLS is to multiply both sides of equation \eqref{eq:statModel} by the constant 
$n\times n$ matrix $P=V^{-1/2}$, in such a way that the variance of the 
transformed error $P\boldsymbol{\epsilon}$ becomes constant amongst the dataset; indeed, 
$\mathrm{\mathbf{Var}}[P\boldsymbol{\epsilon}]=\sigma^2\mathds{1}_n$ since 
$P$ is symmetric. In practice, one may set $\sigma^2=1$ since this constant is the desired variance 
of the transformed variable $P\mathbf{Y}$, in which case  
$V=\mathrm{diag}\bigl\{\widehat{\sigma_1}^2,\widehat{\sigma_2}^2,\dotsc,\widehat{\sigma_n}^2\bigr\}$. The expressions for $\boldsymbol{\widehat{\beta}}$, 
$\mathrm{\mathbf{Var}}\bigl[\boldsymbol{\widehat{\beta}}\bigr]$ and 
$\mathrm{\mathbf{Var}}\bigl[\widehat{\mathbf{Y}}\bigr]$ in the WLS method are obtained by 
replacing $X$ with $PX$ and $\mathbf{Y}$ with $P\mathbf{Y}$ in the usual OLS expressions. 
The WLS expressions are then
\begin{equation}
 \begin{split}
  \label{eq:WLS}
  \boldsymbol{\widehat{\beta}}						&=	\left(X^\intercal V^{-1}X\right)^{-1}X^\intercal V^{-1}\mathbf{Y}	\\
  \mathrm{\mathbf{Var}}\bigl[\boldsymbol{\widehat{\beta}}\bigr]	&=	\sigma^2\left(X^\intercal V^{-1}X\right)^{-1}				\\
  \mathrm{\mathbf{Var}}\bigl[\widehat{\mathbf{Y}}\bigr]		&=	\sigma^2X\,\mathrm{\mathbf{Var}}\bigl[\boldsymbol{\widehat{\beta}}\bigr]X^\intercal.
 \end{split}
\end{equation}
Since we are interested in polynomial models, the last expression of \eqref{eq:WLS} may be 
rewritten to yield $\mathrm{\mathbf{Var}}\bigl[\widehat{\mathbf{Y}}\bigr]$ as a function of $x$:
\begin{align}
 \label{eq:varB}
 \notag\mathrm{\mathbf{Var}}\bigl[\widehat{Y}_x\bigr]		&=	\sigma^2\mathbf{x}^\intercal\mathrm{\mathbf{Var}}\bigl[\boldsymbol{\widehat{\beta}}\bigr]\mathbf{x}											\\
										&=	\sigma^2\left(1,x,x^2,\dotsc,x^{p-1}\right)\mathrm{\mathbf{Var}}\bigl[\boldsymbol{\widehat{\beta}}\bigr]\left(1,x,x^2,\dotsc,x^{p-1}\right)^\intercal\equiv E(x),
\end{align}
where $\widehat{Y}_x=\mathbf{x}^\intercal\widehat{\boldsymbol{\beta}}$ is the linear regression. 
Using \eqref{eq:varB}, the $95\%$ confidence limits of $\widehat{Y}_x$ as 
a function of $x$ is 
\begin{equation}
 \label{eq:confIntReg}
 \left[\widehat{Y}_x-\tau_{n-p}(0.95)E(x),\,\widehat{Y}_x+\tau_{n-p}(0.95)E(x)\right],
\end{equation}
with $\tau_{n-p}(0.95)$ as in \eqref{eq:CI}.
\begin{figure}[htb]
 \begin{center}
  \includegraphics[width=0.8\textwidth]{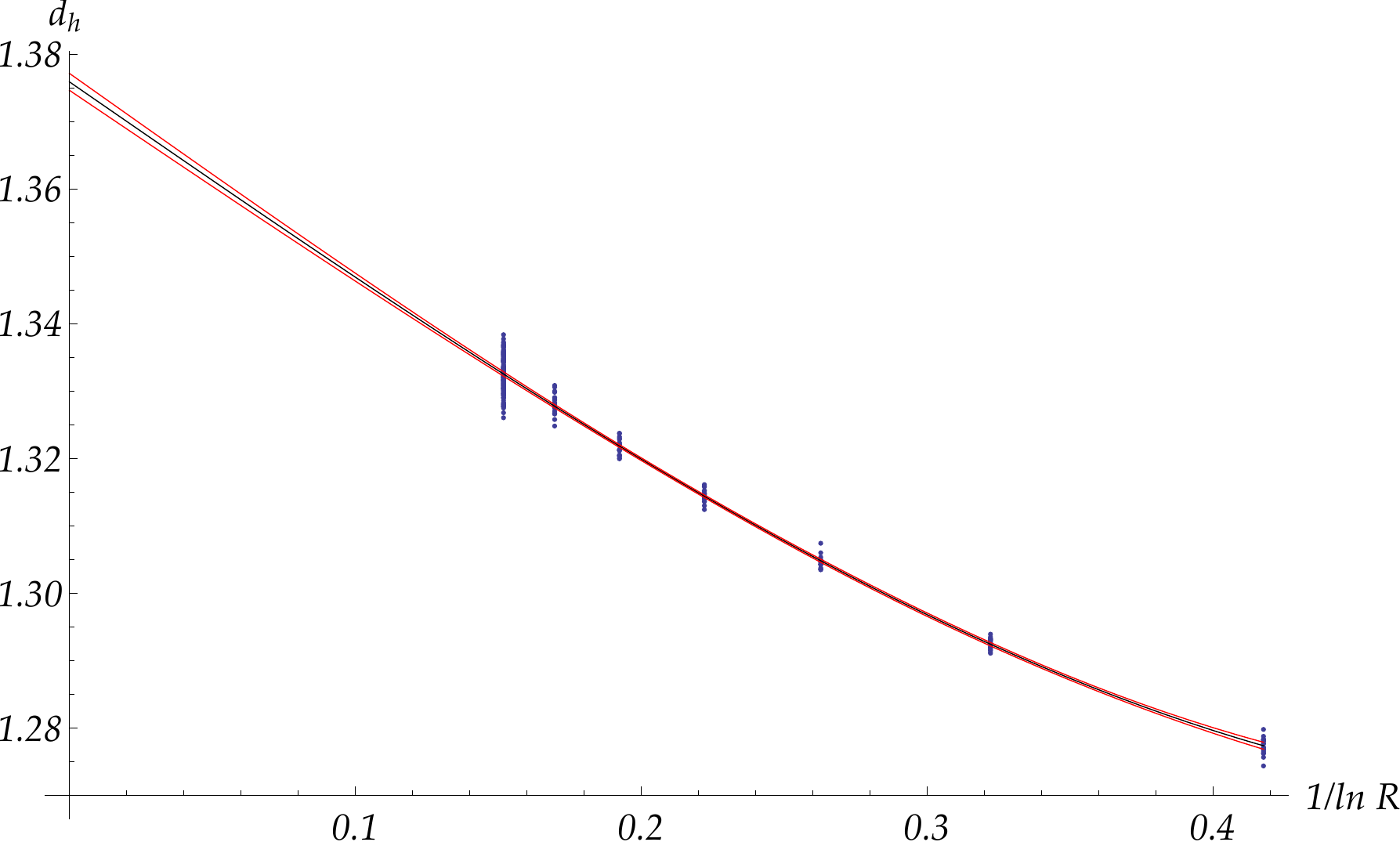}
  \caption[A fitted model.]{\label{fig:fit}The sample points, the linear regression and its $95\%$ 
  confidence limits for $d_h$ of $\mathcal{DLM}(3,4)$ on the $V/H=2$ cylinder.}
 \end{center}
\end{figure}

In figure~\ref{fig:fit}, the results of fitting the model $Y_i=\beta_0+\beta_1X_i+\beta_3X_i^3$ to the 
dataset of the observable $d_h$ (eq. \eqref{eq:hull}) for 
$\mathcal{DLM}(3,4)$ on the $V/H=2$ cylinder are shown. The average of $d_h$ for each run at 
each size $H\times V$ is shown by a small dot. The bottom and top curves are the 
confidence limits obtained by \eqref{eq:confIntReg}, and the central curve corresponds to the fit 
$\widehat{Y}_x$. The $95\%$ confidence interval at $1/ln R=0$ is 
$[1.3744,1.3770]$, thus $\widehat{d_h}=1.3757\pm0.0013$ (cf table \ref{tab:Meas}).

\subsubsection{Model testing}\label{sec:modelTesting}

Once a linear regression has been obtained, it is necessary to attest its validity and quality. To do 
this, the popular Pearson correlation coefficient
$\left.\sum_{i=1}^n\bigl(\widehat{Y}_i-\overline{Y}\bigr)^2\right/\sum_{i=1}^n\bigl(Y_i-\overline{Y}\bigr)^2$ can be of help, but alone may lead 
to biased results, as discussed in \textsection~11.2 of \citet{DS1998}, especially if an extrapolation 
from the fit is intended. To complement the correlation coefficient, one may use the 
{\em$F$-test} that compares different models of linear regression for $\mathbf{Y}$. To do so, we 
must first fix the \emph{full model} that contains all the $\beta$ parameters 
that are sensible to use. For instance, our linear regressions used a polynomial model of order 3. 
Higher order polynomials would have required data on more lattice sizes. The 
full model is thus
\begin{equation}
 \label{eq:exModel}
 Y_i=\beta_0+\beta_1X_i+\beta_2X_i^2+\beta_3X_i^3+\epsilon_i.
\end{equation}
Different models are then compared with the full model. The $F$-test is used to verify \emph{linear 
hypotheses}. For example, we may want to compare the quality of the 
fit obtained, say for a cubic model like \eqref{eq:exModel} but with \mbox{$\beta_2=0$}, or an even 
model where \mbox{$\beta_1=\beta_3=0$}. We might also want to test 
a hypothesis of the form 
\mbox{$\beta_0+2\beta_1=4$} \emph{and} \mbox{$\beta_0+\beta_1+\beta_3=-1$}, where there are 
now two relations to be satisfied at the same 
time. More generally, a linear hypothesis can be written as
\begin{equation}
 \label{eq:Hypo}
 H_0: R\boldsymbol{\beta}=\mathbf{r},
\end{equation}
where $R$ is an $m\times p$ matrix providing $m$ linear relations amongst the $\beta$'s,
where $q$ of these restrictions are linearly independent. The $m\times1$ vector 
$\mathbf{r}$ contains the constants of the $m$ relations.

To verify such a hypothesis, both the estimator $\widehat{\boldsymbol{\beta_r}}$ of the 
\emph{restricted model}, $\mathbf{Y}=X\boldsymbol{\beta_r}+\boldsymbol{\epsilon}$,
\begin{equation}
 \label{eq:betaRestrict}
 \widehat{\boldsymbol{\beta_r}}=\widehat{\boldsymbol{\beta}}+\bigl(X^\intercal V^{-1}X\bigr)^{-1}R^\intercal\Bigl[R\bigl(X^\intercal V^{-1}X\bigr)^{-1}R^\intercal\Bigr]^{-1}\bigl(\mathbf{r}-R\widehat{\boldsymbol{\beta}}\bigr),
\end{equation}
and the estimator $\widehat{\boldsymbol{\beta}}$ of the full model are computed. Then the 
\emph{residual sum of squares} $\mathrm{SSE}\bigl(\widehat{\boldsymbol{\beta_r}}\bigr)$ and 
$\mathrm{SSE}\bigl(\widehat{\boldsymbol{\beta}}\bigr)$ for both models 
are obtained. Their estimators are defined by
\begin{equation}
 \label{eq:SSE}
 \mathrm{SSE}\bigl(\widehat{\boldsymbol{\beta}}\bigr)=\bigl(\mathbf{Y}-X\widehat{\boldsymbol{\beta}}\bigr)^\intercal\bigl(\mathbf{Y}-X\widehat{\boldsymbol{\beta}}\bigr),
\end{equation}
and similarly for $\widehat{\boldsymbol{\beta_r}}$. Finally, the $F$-test for the hypothesis 
\eqref{eq:Hypo} consists in comparing the ratio
\begin{equation}
 \label{eq:Ftest}
 f_{H_0}=\left(\frac{\mathrm{SSE}\bigl(\widehat{\boldsymbol{\beta_r}}\bigr)-\mathrm{SSE}\bigl(\widehat{\boldsymbol{\beta}}\bigr)}{\mathrm{SSE}\bigl(\widehat{\boldsymbol{\beta}}\bigr)}\right)\left(\frac{n-p}{q}\right),
\end{equation}
with the value $z$ for which \mbox{$\int_0^zF(q,n-p,x)\,\mathrm{d}x=\lambda$}, where 
$\lambda\in[0,1]$ (we chose $\lambda=0.95$ in this work) and the 
\emph{$F$-distribution} is
\begin{equation}
 \label{eq:Fdist}
 F(n_1,n_2,x)=\frac{1}{\mathrm{B}\left(\frac{n_1}{2},\frac{n_2}{2}\right)}\left(\frac{n_1}{n_2}\right)^{n_1/2}x^{\frac{n_1}{2}-1}\left(1+\frac{n_1}{n_2}x\right)^{-(n_1+n_2)/2},
\end{equation}
with $\mathrm{B}(x,y)$ the Beta function. Now, if $f_{H_0}\leq z$ then, with probability $\lambda$, 
the dataset does not provide sufficient proof that $H_0$ has to be rejected. 
On the other hand, if $f_{H_0}>z$, then the dataset shows that the hypothesis $H_0$ is implausible 
and should be rejected. Different scores coming from different hypotheses 
concerning the same full model may be compared with each other, the lowest score being the 
most plausible. We use this test to check whether one or more of the coefficients 
$\beta_i$ could be set to zero. When this is a possibility, we note that the estimated fractal 
dimension obtained from the restricted model is, most of the time, closer to the 
theoretical value than that of the full one, and its confidence interval is also smaller. For the 
dataset of the $d_h^{H\times V}$ of $\mathcal{DLM}(3,4)$, the full cubic model of 
\eqref{eq:exModel} gives $\widehat{d_h}=1.377\pm0.007$, while it is $1.3757\pm0.0013$ for the 
restricted model having the lowest score, that is the one with $\beta_2=0$.


\section{Logarithmic minimal models}\label{sec:LM}

\begin{figure}[htb!]
 \begin{center}
  \includegraphics[width=0.2\textwidth]{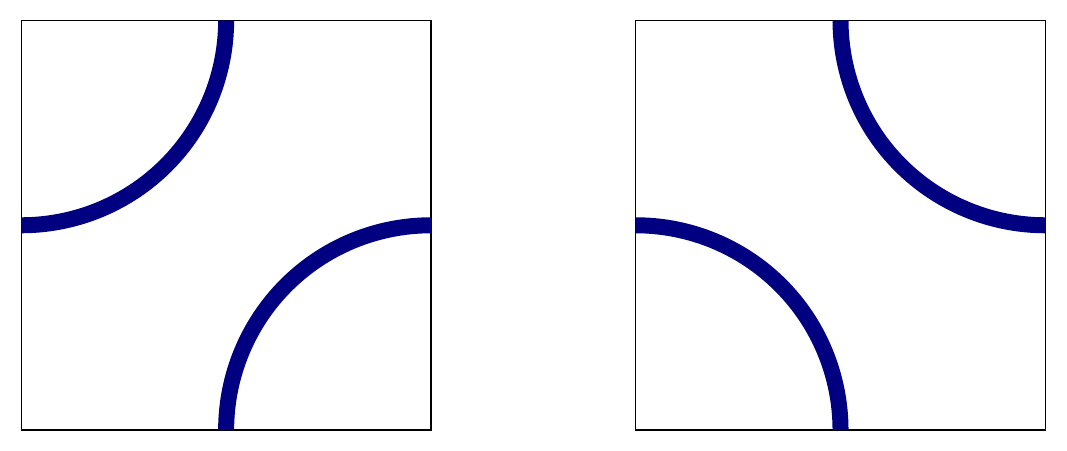}
  \caption[The two faces of $\mathcal{LM}$.]{\label{fig:LM}The two equally weighted faces of $\mathcal{LM}$.}
 \end{center}
\end{figure}
\begin{figure}[htb!]
 \begin{center}
  \includegraphics[width=0.3\textwidth]{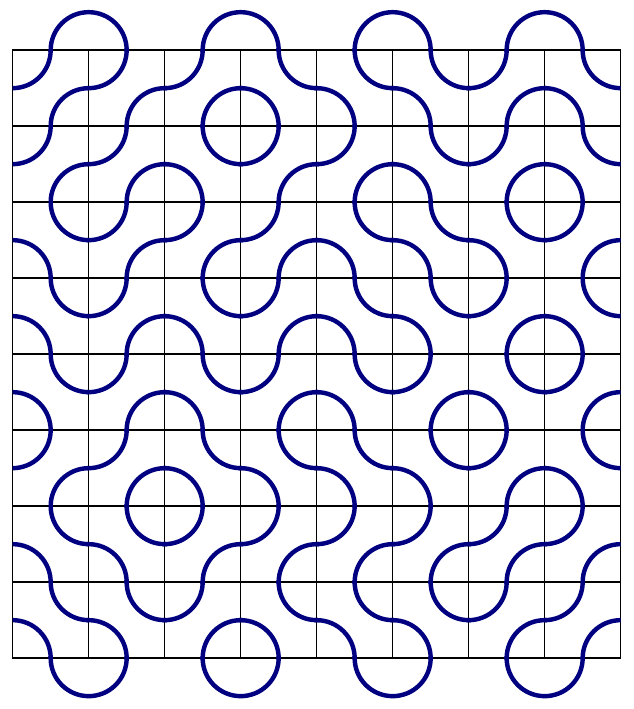}
  \caption[An $\mathcal{LM}$ configuration on the cylinder.]{\label{fig:LMconf}An example of an 
  $8\times8$ configuration on the cylinder.}
 \end{center}
\end{figure}

Here the lattice models used by \citet{PRZ2006} to define the logarithmic minimal 
models $\mathcal{LM}(p,p')$ are recalled. They were used in 
\citep{Saint-Aubin2009} for simulations of the dense models. The underlying loop gas is based on two 
elementary faces, illustrated in figure~\ref{fig:LM}. At the isotropic point, the two 
faces are given an equal weight $\rho_1=\rho_2$ which can be fixed to $1$. The partition function 
of the loop gas is then
\begin{equation}
 \label{eq:partFuncLM}
 Z=\sum_{\mathcal{L}}\beta^N,
\end{equation}
where $\mathcal{L}$ is the set of all possible loop configurations, $\beta$ the loop fugacity, 
and $N$ the number of loops in a given configuration. These models do not 
correspond to a particular solution of the $\mathcal{DLM}$ models (\eqref{eq:OnSquareLoop} 
and \eqref{eq:YBsol}) as no value of $\lambda\in[-\pi/2,\pi/2]$ in any of the 
critical regimes found in \citet{Blote1989} yields $u=v=0$, $w\neq0$ and no empty face. The 
simulations in \citep{Saint-Aubin2009} were done on a cylinder. The boundary 
conditions consisted in half-circles added at the extremities of the cylinder, as shown in 
figure~\ref{fig:LMconf}. Note that, with these boundary conditions, there are 
$|\mathcal{L}|=2^{H\times V}$ possible configurations.

It is these lattice models whose continuum scaling limit was given the name of logarithmic minimal 
models $\mathcal{LM}(p,p')$. Their loop fugacity is given by 
$\beta=-2cos\left(\frac{\pi}{\bar{\kappa}}\right)$ with $\bar{\kappa}=p'/p$ and they are believed to 
be described by logarithmic CFTs whose central charge and conformal weights are 
given by \eqref{eq:centCharge} and \eqref{eq:confWeights}.


\newpage
\bibliographystyle{myplainnat} 
\bibliography{u11}

\begin{thebibliography}{28}
\providecommand{\natexlab}[1]{#1}
\providecommand{\url}[1]{\texttt{#1}}
\expandafter\ifx\csname urlstyle\endcsname\relax
  \providecommand{\doi}[1]{doi: #1}\else
  \providecommand{\doi}{doi: \begingroup \urlstyle{rm}\Url}\fi

\bibitem[Aharony and Asikainen(2003)]{Aharony2002}
A.~Aharony and J.~Asikainen.
\newblock \emph{Fractal dimensions and corrections to scaling for critical
  {P}otts clusters}.
\newblock Fractals, {\bf 11}:\penalty0 3--7, 2003,
  \href{http://arxiv.org/abs/cond-mat/0206367}{\tt arXiv:cond-mat/0206367}.

\bibitem[Arguin(2002)]{Arguin2002}
L.-P. Arguin.
\newblock \emph{Homology of {F}ortuin-{K}asteleyn clusters of {P}otts models on
  the torus}.
\newblock J. Stat. Phys., {\bf 109}:\penalty0 301--310, 2002,
  \href{http://arxiv.org/abs/hep-th/0111193}{\tt arXiv:hep-th/0111193}.

\bibitem[Asikainen et~al.(2003)Asikainen, Aharony, Mandelbrot, Rauch, and
  Hovi]{Asikainen2003}
J.~Asikainen, A.~Aharony, B.~B. Mandelbrot, E.~M. Rauch, and J.~P. Hovi.
\newblock \emph{Fractal geometry of critical {P}otts clusters}.
\newblock Eur. Phys. Jour. B., {\bf 34}:\penalty0 479--487, 2003,
  \href{http://arxiv.org/abs/cond-mat/0212216}{\tt arXiv:cond-mat/0212216}.

\bibitem[Beffara(2008)]{Beffara2008}
V.~Beffara.
\newblock \emph{The dimension of the {SLE} curves}.
\newblock Ann. Probab, {\bf 36\penalty0 (4)}:\penalty0 1421--1452, 2008,
  \href{http://arxiv.org/abs/math/0211322}{\tt arXiv:math/0211322v3 [math.PR]}.

\bibitem[Blöte and Nienhuis(1989)]{Blote1989}
H.~W.~J. Blöte and B.~Nienhuis.
\newblock \emph{Critical behaviour and conformal anomaly of the
  $\mathcal{O}(n)$ model on the square lattice}.
\newblock J. Phys. A: Math. Gen., {\bf 22}:\penalty0 1415--1438, 1989.

\bibitem[Camia and Newman(2006)]{Camia2006}
F.~Camia and C.~M. Newman.
\newblock \emph{Two-dimensional critical percolation: the full scaling limit}.
\newblock Commun. Math. Phys., {\bf 268}:\penalty0 1--38, 2006,
  \href{http://lanl.arxiv.org/abs/math/0605035}{\tt arXiv:math/0605035v1
  [math.PR]}.

\bibitem[Chayes and Machta(1998)]{Chayes1998}
L.~Chayes and J.~Machta.
\newblock \emph{Graphical representations and cluster algorithms {II}}.
\newblock Physica A, {\bf 254}:\penalty0 477--516, 1998.

\bibitem[Deng et~al.(2007)Deng, Garoni, Guo, Blöte, and Sokal]{Deng2007}
Y.~Deng, T.~M. Garoni, W.~Guo, H.~W.~J. Blöte, and A.~D. Sokal.
\newblock \emph{Cluster simulations of loop models on two-dimensional
  lattices}.
\newblock Phys. Rev. Lett., {\bf 98}:\penalty0 120601, 2007,
  \href{http://arxiv.org/abs/cond-mat/0608447v3}{\tt arXiv:cond-mat/0608447v3
  [cond-mat.stat-mech]}.

\bibitem[Draper and Smith(1998)]{DS1998}
N.~R. Draper and H.~Smith.
\newblock \emph{Applied Regression Analysis}.
\newblock New York, 3rd edition, 1998.

\bibitem[Dubail et~al.(2010)Dubail, Jacobsen, and Saleur]{Dubail2009}
J.~Dubail, J.~L. Jacobsen, and H.~Saleur.
\newblock \emph{Conformal boundary conditions in the critical $\mathcal{O}(n)$
  model and dilute loop models}.
\newblock Nucl. Phys. B, {\bf 827}:\penalty0 457--502, 2010,
  \href{http://arxiv.org/abs/0905.1382}{\tt arXiv:0905.1382v1}.

\bibitem[Duplantier(1987)]{Duplantier1987}
B.~Duplantier.
\newblock \emph{Critical exponents of {M}anhattan {H}amiltonian walks in two
  dimensions, from {P}otts and $\mathcal{O}(n)$ models}.
\newblock J. Stat. Phys., {\bf 49}:\penalty0 411--431, 1987.

\bibitem[Duplantier(1989)]{Duplantier1989}
B.~Duplantier.
\newblock \emph{Two-dimensional fractal geometry, critical phenomena and
  conformal invariance}.
\newblock Phys. Rep., {\bf 184\penalty0 (2--4)}:\penalty0 229--257, 1989.

\bibitem[Duplantier(2000)]{Duplantier2000}
B.~Duplantier.
\newblock \emph{Conformally invariant fractals and potential theory}.
\newblock Phys. Rev. Lett., {\bf 84\penalty0 (7)}:\penalty0 1363--1367, 2000.

\bibitem[Fishman(2006)]{Fishman2006}
G.~S. Fishman.
\newblock \emph{A First Course in Monte Carlo}.
\newblock Belmont, 2006.

\bibitem[Grossman and Aharony(1986)]{Grossman1986}
T.~Grossman and A.~Aharony.
\newblock \emph{Structure and perimeters of percolation clusters}.
\newblock J. Phys. A: Math. Gen., {\bf 19}:\penalty0 L745--L751, 1986.

\bibitem[Kager and Nienhuis(2004)]{Kager2004}
W.~Kager and B.~Nienhuis.
\newblock \emph{A guide to stochastic {L}oewner evolution and its
  applications}.
\newblock J. Stat. Phys., {\bf 115}:\penalty0 1149--1229, 2004,
  \href{http://arxiv.org/abs/math-ph/0312056}{\tt arXiv:math-ph/0312056v3}.

\bibitem[Langlands et~al.(2000)Langlands, Lewis, and
  Saint-Aubin]{Langlands2000}
R.~Langlands, M.-A. Lewis, and Y.~Saint-Aubin.
\newblock \emph{Universality and conformal invariance for the {I}sing model in
  domains with boundary}.
\newblock J. Stat. Phys., {\bf 98}:\penalty0 131--244, 2000.

\bibitem[Mandelbrot(1990)]{Mandelbrot1990}
B.~B. Mandelbrot.
\newblock \emph{Negative fractal dimensions and multifractals}.
\newblock Physica A, {\bf 163}:\penalty0 306--315, 1990.

\bibitem[Nienhuis(1990)]{Nienhuis1990}
B.~Nienhuis.
\newblock \emph{Critical and multicritical $\mathcal{O}(n)$ models}.
\newblock Physica A, {\bf 163}:\penalty0 152--157, 1990.

\bibitem[Pearce et~al.(2006)Pearce, Rasmussen, and Zuber]{PRZ2006}
P.~A. Pearce, J.~Rasmussen, and J.-B. Zuber.
\newblock \emph{Logarithmic minimal models}.
\newblock J. Stat. Mech., P11017, 2006,
  \href{http://arxiv.org/abs/hep-th/0607232}{\tt arXiv:hep-th/0607232v3}.

\bibitem[Pinson(1994)]{Pinson1994}
T.~H. Pinson.
\newblock \emph{Critical percolation on the torus}.
\newblock J. Stat. Phys., {\bf 75}:\penalty0 1167--1177, 1994.

\bibitem[Rohde and Schramm(2005)]{Rohde2005}
S.~Rohde and O.~Schramm.
\newblock \emph{Basic properties of {SLE}}.
\newblock Ann. Math., {\bf 161}:\penalty0 883--924, 2005,
  \href{http://arxiv.org/abs/math/0106036}{\tt arXiv:math/0106036v4 [math.PR]}.

\bibitem[Saint-Aubin et~al.(2009)Saint-Aubin, Pearce, and
  Rasmussen]{Saint-Aubin2009}
Y.~Saint-Aubin, P.~A. Pearce, and J.~Rasmussen.
\newblock \emph{Geometric exponents, {SLE} and logarithmic minimal models}.
\newblock J. Stat. Mech., P02028, 2009,
  \href{http://arxiv.org/abs/arXiv:0809.4806}{\tt arXiv:0809.4806v2}.

\bibitem[Saleur and Duplantier(1987)]{Saleur1987}
H.~Saleur and B.~Duplantier.
\newblock \emph{Exact determination of the percolation hull exponent in two
  dimensions}.
\newblock Phys. Rev. Lett., {\bf 58}:\penalty0 2325--2328, 1987.

\bibitem[Sheffield(2006)]{Sheffield2006}
S.~Sheffield.
\newblock \emph{Exploration trees and conformal loop ensembles}.
\newblock Duke Math. J., {\bf 147\penalty0 (1)}:\penalty0 79--129, 2006,
  \href{http://arxiv.org/abs/math.PR/0609167}{\tt arXiv:math/0609167v2
  [math.PR]}.

\bibitem[Stanley(1977)]{Stanley1977}
H.~E. Stanley.
\newblock \emph{Cluster shapes at the percolation threshold: an effective
  cluster dimensionality and its connection with critical-point exponents}.
\newblock J. Phys. A: Math. Gen., {\bf 10}:\penalty0 L211--L220, 1977.

\bibitem[Swendsen and Wang(1987)]{Swendsen1987}
R.~H. Swendsen and J.-S. Wang.
\newblock \emph{Nonuniversal critical dynamics in {M}onte {C}arlo simulations}.
\newblock Phys. Rev. Lett., {\bf 58\penalty0 (2)}:\penalty0 86--88, 1987.

\bibitem[Werner(2008)]{Werner2005}
W.~Werner.
\newblock \emph{The conformally invariant measure on self-avoiding loops}.
\newblock J. Amer. Math. Soc., {\bf 21}:\penalty0 137--169, 2008,
  \href{http://lanl.arxiv.org/abs/math/0511605}{\tt arXiv:math/0511605v3
  [math.PR]}.

\end{thebibliography}

\end{document}